\documentclass[10pt]{article}
\usepackage[a4paper, total={6.5in, 8in}]{geometry}

\usepackage[T1]{fontenc}
\usepackage{lmodern}

\usepackage{fancyhdr}
\usepackage{graphicx}
\usepackage{pifont}
\usepackage{dcolumn}
\usepackage{bm}
\usepackage{caption,subcaption}

\usepackage{amsthm}
\usepackage{latexsym}
\usepackage{graphics}
\usepackage{amssymb}
\usepackage{amsfonts}
\usepackage{mathrsfs}
\usepackage{bm}
\usepackage{amsmath}
\usepackage{color}
\usepackage[dvipsnames]{xcolor}
\usepackage{arydshln} 
\usepackage[title]{appendix}

\usepackage{authblk}

\usepackage{cancel}

\RequirePackage{graphicx}
\RequirePackage{flushend}
\RequirePackage[colorlinks,citecolor=blue,urlcolor=blue,linkcolor=blue]{hyperref}

\usepackage{mathtools}  

\newtheorem{theorem}{Theorem}[section]

\newtheorem{lemma}[theorem]{Lemma}

\newtheorem{definition}{Definition}[section]
\newcommand{\oh}{\dfrac{1}{2}}

\newcommand{\rs}{\text{Res}\,}
\newcommand{\gm}{g_{-}}
\newcommand{\gp}{g_{+}}
\newcommand{\gpm}{g_{\pm}}
\newcommand{\dz}{\mathrm{d}z}
\newcommand{\dpp}{\mathrm{d}p}
\newcommand{\du}{\mathrm{d}u}
\newcommand{\dtau}{\mathrm{d}\tau}

\begin{document}


\title{Casimir energy for spinor fields with $\delta$-shell potentials}
\bigskip

\author[1]{Guglielmo Fucci\footnote{e-mail: fuccig@ecu.edu}}
\author[2]{C\'esar Romaniega\footnote{e-mail: cesar.romaniega@uva.es (corresponding author)}}

\affil[1]{\small{Department of Mathematics, East Carolina University}\\ \small{Greenville, North Carolina 27858, USA.}}
\affil[2]{\small{Departamento de F\'isica Te\'orica, At\'omica y \'Optica, Universidad de Valladolid}\\\small{Valladolid, 47011, Spain.}}
\bigskip

\bigskip


\maketitle

\begin{abstract}
This work analyzes the Casimir energy of a massive spinor field propagating in flat space endowed with a spherically 
symmetric $\delta$-function potential. By utilizing the spectral zeta function regularization method, the Casimir energy is evaluated 
after performing a suitable analytic continuation. Explicit numerical results are provided for specific cases in which the Casimir energy is 
unambiguously defined. The results described in this work represent a generalization of the MIT bag model for spinor fields.

\end{abstract}
\section{Introduction}\label{sec:I}

The Casimir effect, first theorized by H. B. G. Casimir in his famous 1948 paper
\cite{casimir1948attraction}, is one of the most studied phenomena in the ambit of quantum
field theory under the influence of external fields and boundary conditions. Since the early
1990's, when interest in the Casimir effect started to truly gain momentum, calculations of
the Casimir energy and force were performed for a large number of geometric configurations
and boundary conditions (see e.g. the monographs \cite{bordag2001new,milton2004casimir} and
references therein). The majority of these investigations, however, have been
focused on the Casimir energy and force for systems consisting of both massive and massless scalar fields. Spinor fields, on the other hand, have not enjoyed 
the same level of research interest. The main reason for this apparent lack of interest
resides, most likely, in the fact that the formalism needed to treat spinor fields is more 
involved than the one used to study scalar fields. In fact, the propagation of a quantum field is 
modeled by an appropriate operator (e.g. a first order Dirac operator for spinor fields and a second order Laplace operator 
for scalar fields) endowed with a set of boundary conditions (or integrability conditions at infinity in the unbounded case)
which renders the problem self-adjoint. While self-adjoint extensions for Laplace-type operators
are very well studied, those for Dirac operators are more nuanced and less investigated.

One type of boundary condition that is most often used, as it makes the Dirac operator self-adjoint in a bounded region of space, is the MIT bag boundary condition \cite{johnson1975,arriza2017}. 
This boundary condition essentially ensures that the spinor field, while free to propagate inside a given region, cannot escape through its boundary. The MIT bag boundary condition was introduced in order to construct a model of quark and gluon confinement in hadrons \cite{chodos1974new,chodos1974,chodos1975}. Since its introduction, the MIT bag has become, effectively, the only physically relevant model of fermion confinement. In fact, to the best of our knowledge, models of spinors propagating within a region with a spherical boundary have considered exclusively the MIT bag boundary condition.

Imposing the MIT bag boundary condition on a Dirac operator acting on spinor fields propagating in a bounded domain is 
not the only way of obtaining a self-adjoint system. In fact, self-adjoint extensions of the Dirac operator can 
be obtained by introducing point-supported potentials. This research direction started with the study of the self-adjointness of the Dirac operator with  point-supported potentials in one-dimension \cite{albeverio2005solvable,gesztesy1987}
and continued with an analogous investigation where the potential has support on a spherical shell in $\mathbb{R}^{3}$ \cite{dittrich1989dirac}. More recently, self-adjoint extensions of the Dirac operator have been analyzed for potentials 
supported on the boundary of an arbitrary region of $\mathbb{R}^{3}$ \cite{arriza2014,arriza2015,behrndt2019dirac}. Point supported 
potentials are of particular interest especially because they can be used to model boundary conditions. 
In fact, potentials defined on $\delta$-shells which allow for a complete separation between the interior and the exterior 
regions can be used as alternative models of confinement to the MIT bag. Moreover, there exist a class of $\delta$-shell potentials 
that permits an interaction between the interior and exterior regions. The latter potentials could be used to 
describe a \emph{soft} bag in which the interior spinors can escape the $\delta$-shell region. These models could be 
of great interest since there exists evidence for quark tunneling between close nucleons (see e.g. \cite{goldman1984} and references therein). 

It is the goal of this work to analyze the Casimir effect for massive spinor fields propagating in $\mathbb{R}^{3}$
under the influence of a spherical $\delta$-shell potential. More specifically, we consider the Dirac operator coupled with a linear combination of electrostatic and Lorentz scalar potentials supported on a spherical $\delta$-shell. 
While it is possible, in principle, to investigate 
the more general case of potentials supported on the boundary of an arbitrary region of $\mathbb{R}^{3}$, we choose to focus on the spherical $\delta$-shell in order to present an exactly solvable case and to deal with a larger set of self-adjoint extensions. The Casimir effect for massive fermionic fields was analyzed in detail in \cite{elizalde1998casimir} by using the zeta function regularization method. In that paper, the authors considered a massive spinor field endowed with the MIT bag boundary condition on a spherical surface.
Our paper serves as an extension of the work performed in \cite{elizalde1998casimir}. 
In fact by considering spinor fields in $\mathbb{R}^{3}$ under the presence of a spherical 
$\delta$-shell potential, we are able to compute the Casimir energy for a variety of boundary conditions which are modeled by selecting particular linear combinations of the electrostatic and Lorentz scalar $\delta$-shell potentials. In is important to mention that we can reproduce the results obtained when the MIT bag boundary condition is imposed by simply choosing specific values for the parameters of the linear combination that characterizes 
the $\delta$-shell potential.

In order to evaluate the Casimir energy for spinor fields under the influence of a 
$\delta$-shell spherical potential, we utilize the spectral zeta function 
regularization method. We focus, in this work, on those $\delta$-shell potentials that
lead to spinor confinement since they represent a generalization of the MIT bag model. 
The outline of the paper is as follows.
In the next Section, we outline the mathematical framework of our spinor system and define its spectral zeta function. Subsequently, we obtain an integral representation for the spectral zeta function and perform a suitable analytic continuation to a wider region of the complex plane. We then utilize the analytically continued expressions to analyze the 
Casimir energy in the massless case. The conclusions, then, summarize the
main results of our paper and point to a few directions for future research.    
We would like to mention at this point that, as we will see in the next Section, the evaluation of the spectral zeta function involves the analysis of the characteristic function which implicitly provides the eigenvalues of the system. The characteristic function is also known, in the literature, as the mode-generating function.

\section{Dirac operator and spectral zeta function}
We consider a spinor field of mass $m$ propagating in $\mathbb{R}^{3}$ under the influence of a spherically symmetric $\delta$-shell interaction. The starting point of our analysis
is the free Dirac Hamiltonian acting on functions in the Hilbert space $L^{2}(\mathbb{R}^{3})\otimes \mathbb{C}^{4}$  
\begin{equation}\label{eq:DiracFree}
H_0 = -i \alpha_i \partial_i + m\beta\;,
\end{equation}
where $\alpha_i$ and $\beta$ represent the Dirac matrices
\begin{equation}
 \beta = \left(
\begin{array}{cc}
	I & 0 \\
	0 & -I \\
\end{array}
\right),
\quad \alpha_{i}=\left(
\begin{array}{cc}
	0 & \sigma _k \\
	\sigma _k & 0 \\
\end{array}
\right),
\end{equation}
with $\sigma _k$ being the Pauli matrices and $I$ and $0$ denoting the $2\times 2$ identity and zero matrices, respectively. Among all possible self-adjoint extensions, we only consider those that are rotationally and space-reflection invariant.  This requirement 
leads naturally to a decomposition of the Dirac Hamiltonian into an angular and a radial part.
In fact, the state Hilbert space can be decomposed into an orthogonal sum of subspaces referring to the total angular momentum, its third component, and the parity as follows \cite{dittrich1989dirac} 
\begin{equation}\label{eq:parametrization}
	\begin{aligned}
	\mathcal{H}&=\displaystyle\bigoplus^{\infty} _{j=1/2}\ \ \bigoplus^{j+1/2} _{\ell=j-1/2}\ \ \bigoplus^{j}_{M=-j}\mathcal{H}_{j \ell M}\\[0.5ex]
	\mathcal{H}_{j \ell M}&=\left\{\varphi \in \mathcal{H}:\varphi(\vec{r})=\frac{1}{r}\left(
	\begin{array}{c}
		f(r) \mathcal{Y}^{+\sigma }_{\ell M}(\theta ,\phi ) \\
	i	g(r) \mathcal{Y}^{-\sigma }_{\ell'M}(\theta ,\phi ) \\
	\end{array}
	\right);f,g\in L^2\left(\mathbb{R}_+,\text{d}r \right)\right\},
	\end{aligned}
\end{equation}
where we have introduced the spherical spinors
\begin{equation}
	\mathcal{Y}^{\sigma }_{\ell M}=\left(
	\begin{array}{c}
		\sigma  \sqrt{\dfrac{2 \ell+2 M \sigma +1}{2 (2 \ell+1)}} Y_\ell^{M-\frac{1}{2}}(\theta ,\phi ) \\[0.5ex]
		\sqrt{\dfrac{2 \ell-2 M \sigma +1}{2 (2 \ell+1)}} Y_\ell^{M+\frac{1}{2}}(\theta ,\phi ) \\
	\end{array}
	\right)\;.
\end{equation}
In the last formula $\sigma=\pm 1$ describes the spin projection, $\ell'=\ell+\sigma$ and $Y_l^{M}(\theta ,\phi )$ denotes the spherical harmonics. 
We would like to point out that the decomposition in (\ref{eq:parametrization}) is equivalent to the one provided in \cite{dittrich1989dirac}
with the only difference stemming from the fact that we have chosen the radial functions $f(r)$ and $g(r)$ to be
in $L^2\left(\mathbb{R}_+,\text{d}r \right)$ (in accordance with the monograph \cite{grant2007relativistic}) while the radial functions in 
\cite{dittrich1989dirac} belong to $L^2\left(\mathbb{R}_+,r^2\text{d}r \right)$. In this setting, the Dirac Hamiltonian has an analogous decomposition 
\begin{equation}\label{hamiltonian}
H_0=\displaystyle\bigoplus^{\infty} _{j=1/2}\ \ \bigoplus^{j+1/2} _{\ell=j-1/2}\ \ \bigoplus^{j}_{M=-j}H_{j \ell M}\;.
\end{equation}
The radial part of the operator $H_{j \ell M}$ can then be separated out in each subspace 
$\mathcal{H}_{j \ell M}$ as proved in \cite{dittrich1989dirac}. This fact leads to the important assertion that the self-adjoint extensions of the Dirac Hamiltonian $H_0$ can be obtained from those of its radial part.
The latter, then, can be explicitly found by rewriting (\ref{eq:DiracFree}) in spherical coordinates. By performing this procedure, which is well-known and is described, for example, in \cite{grant2007relativistic}, the eigenstates of the equation
\begin{equation}\label{0}
H_0\varphi = E\varphi\;,
\end{equation}
can be written as in (\ref{eq:parametrization}) with the radial functions satisfying the following
set of coupled first order differential equations
\begin{equation}\label{eq:System1}
	\begin{aligned}
& \left(-\frac{\textrm{d}}{\textrm{d}r}+\dfrac{\kappa }{r}\right)g(r)+mf(r)=E f(r),\\[0.5ex]
& \left(+\frac{\textrm{d}}{\textrm{d}r}+\dfrac{\kappa }{r}\right)f(r)-m g(r)=E g(r),
	\end{aligned}
\end{equation}
which can be decoupled into two ordinary second order differential equations
\begin{eqnarray}\label{1}
\left(\frac{\textrm{d}^2}{\textrm{d}r^2}-\frac{\kappa(\kappa+1)}{r^2}+p^2\right)f(r)=0\;,\newline\\ 
\left(\frac{\textrm{d}^2}{\textrm{d}r^2}-\frac{\kappa(\kappa-1)}{r^2}+p^2\right)g(r)=0\;,
\end{eqnarray}
where the index $\kappa=-\sigma(j+1/2)$ when $\ell=j-\sigma/2$ \cite{grant2007relativistic} and $p=\sqrt{E^2-m^2}$.
The differential equations in (\ref{1}) are of the Bessel type and their general solutions are proportional to cylinder functions $C_{\nu}(r)$ (see e.g. \cite{olver2010nist}, Chapter 10). From these last remarks, one can then conclude that the eigenfunctions of (\ref{0}) are \cite{grant2007relativistic} 
\begin{eqnarray}\label{2}
\varphi^{\sigma}(\textbf{r})=\frac{1}{\sqrt{ {r} }}\left(
	\begin{array}{c}
	i	A_1 C_{j-\frac{\sigma-1}{2}}(p r) \mathcal{Y}^{+\sigma }_{\ell M}(\theta ,\phi ) \\
		A_2 C_{j+\frac{\sigma+1}{2}}(p r) \mathcal{Y}^{-\sigma }_{\ell'M}(\theta ,\phi ) \\
	\end{array}
	\right)\;.
\end{eqnarray}
The arbitrary coefficients $A_1$ and $A_2$ are not independent since the radial functions are coupled according to (\ref{eq:System1}).
By substituting the cylinder function solutions into (\ref{eq:System1}) we find
\begin{eqnarray}\label{3}
\varphi^{\sigma}(\textbf{r})=\frac{\mathcal{A}}{\sqrt{{r}}}\left(
	\begin{array}{c}
	i	 C_{j-\frac{\sigma-1}{2}}(p r) \mathcal{Y}^{+\sigma }_{\ell M}(\theta ,\phi ) \\
		-\sigma\sqrt{\frac{E-m}{E+m}} C_{j+\frac{\sigma+1}{2}}(p r) \mathcal{Y}^{-\sigma }_{\ell'M}(\theta ,\phi ) \\
	\end{array}
	\right)\;,
\end{eqnarray}
where $\mathcal{A}$ represents an overall normalization constant. As it was proved in \cite{dittrich1989dirac}, any self-adjoint extension of \eqref{hamiltonian} is characterized by  boundary conditions imposed on spinors $\phi\in L^{2}(\mathbb{R_+})\otimes \mathbb{C}^{2}$, which are absolutely continuous on $(0,R)$ and $(R,\infty)$, of the form
\begin{eqnarray}\label{BC1}
\phi(R_-)=e^{i\alpha} A \phi(R_+)\;,
\end{eqnarray}  
where $\alpha\in[0,\pi)$ and $A$ is a real $2\times 2$ matrix with $\det A=1$, or
\begin{equation}\label{BC2}
\left(
	\begin{array}{cc}
		c_1 & c_2 \\
		0 & 0 \\
	\end{array}
	\right)\phi(R_-)+\left(
	\begin{array}{cc}
		0 & 0 \\
		d_1 & d_2 \\
	\end{array}
	\right)\phi(R_+)=0\;,
\end{equation}
where the coefficients in the above matrices are real and both matrices are not zero \cite{dittrich1989dirac}. In summary, any self-adjoint extension of $H_0$ can be identified 
by the particular boundary condition, either of the type (\ref{BC1}) or of the type (\ref{BC2}) imposed on the radial part of the solutions (\ref{3}) that are in $L^{2}(\mathbb{R_+})\otimes \mathbb{C}^{2}$ and are absolutely continuous on $(0,R)$ and $(R,\infty)$. In order to proceed with the analysis, it is necessary to introduce, at this point, the potential.   
\begin{definition}[$\delta$-shell potential]
Let $g_s$ and $g_v$ be real (coupling) constants, the $\delta$-shell potential is defined as
\begin{equation}\label{eq:Potential}
V(r)=g_s\beta\delta(r-R)+g_v I_{4}\delta(r-R)\;.
\end{equation}
\end{definition}
This potential is constructed as a linear combination of a Lorentz scalar potential (with coupling constant $g_s$) and an electrostatic $\delta$-shell potential (with coupling constant $g_v$). 
The self-adjoint extensions of the operator $H_0$ endowed with the $\delta$-shell potential can be obtained directly from 
the self-adjoint extensions of the free operator $H_0$ described above. This occurs because
the $\delta$-shell potential acts as a boundary positioned at $R$ with associated boundary conditions which are obtained by integrating the eigenvalue equation 
\begin{equation}\label{eq:eigenvalueeq}
(H_0+V(r))\varphi=E\varphi
\end{equation}  
over the interval $(R-\varepsilon,R+\varepsilon)$ and by then taking the limit $\varepsilon\to 0^+$. This process leads, for the $\delta$-shell potential, to the following boundary, or matching, conditions (cf. \cite{dittrich1989dirac})
\begin{equation}\label{matching}
\left[i\tau+\frac{1}{2}\mathcal{G}\right]\phi^{\sigma}(R_-)+
\left[-i\tau+\frac{1}{2}\mathcal{G}\right]\phi^{\sigma}(R_+)=0\;,
\end{equation} 
where
\begin{equation}\label{matrixG}
\tau=\left(
	\begin{array}{cc}
		0 & 1 \\
		1 & 0 \\
	\end{array}
	\right)\;,\quad\textrm{and}\quad \mathcal{G}=\left(
	\begin{array}{cc}
		\gp & 0 \\
		0 & \gm \\
	\end{array}
	\right)\;,\quad \gpm=g_v\pm g_s\;,
\end{equation}
and $\phi^{\sigma}$ represents just the radial part of $\varphi^{\sigma}$ in (\ref{3}), that is,
\begin{equation}\label{radialpart}
\phi^{\sigma}(r)=\frac{1}{\sqrt{r}}\left(
	\begin{array}{c}
	i	 C_{j-\frac{\sigma-1}{2}}(p r)  \\
		-\sigma\sqrt{\frac{E_-}{E_+}} C_{j+\frac{\sigma+1}{2}}(p r) \\
	\end{array}
	\right)\;,\quad E_{\pm}=E\pm m\;.
\end{equation}
The important point, now, is that the matching conditions (\ref{matching}) generated by the $\delta$-shell potential are either of the form (\ref{BC1}) or of the form (\ref{BC2}) which implies that they characterize any self-adjoint extension of the Hamiltonian $H_0$ endowed with the potential $V(r)$.
We can therefore divide the self-adjoint extensions of $H_0+V(r)$ into two distinct classes:
\begin{itemize}
\item \emph{Non-Confining}. \newline
This class contains all the self-adjoint extensions characterized by the boundary conditions (\ref{matching}) and satisfying 
\begin{equation}
\det\left[-i\tau+\frac{1}{2}\mathcal{G}\right]=\gp\gm+4\neq 0\;.
\end{equation}
In this case the matrix $\left[-i\tau+\frac{1}{2}\mathcal{G}\right]$ is invertible and the matching conditions (\ref{matching}) can be expressed as
\begin{equation}
\phi^{\sigma}(R_+)=\frac{1}{\gp\gm+4}\left(
	\begin{array}{cc}
		4-\gp\gm & -4 i \gm \\
		-4 i \gp & 4-\gp\gm \\
	\end{array}
	\right)\phi^{\sigma}(R_-)\;,
\end{equation}
which share the same form as those in (\ref{BC1}). These types of boundary conditions allow the spinors propagating inside the spherical shell to interact with those outside the spherical shell. In other words, in this case the potential $V(r)$ models a 
\emph{soft wall}.

\item \emph{Confining}.\newline
This class consists of all the self-adjoint extensions characterized by the boundary conditions (\ref{matching}) and satisfying, instead,  
\begin{equation}\label{constraint}
\det\left[-i\tau+\frac{1}{2}\mathcal{G}\right]=\gp\gm+4=0\;.
\end{equation}
By multiplying (\ref{matching}) by the matrix $(i\tau+1/2\tau\mathcal{G}\tau)$  and by noticing that 
$\tau^2=I$ and $(\tau\mathcal{G})^{2}=-4I$, we obtain the following matching conditions for the interior of the spherical region
\begin{equation}\label{eq:matchint}
\left[I-\frac{1}{2}\tau\mathcal{G}\right]\phi^{\sigma}(R_-)=0\;.
\end{equation}
On the other hand, by multiplying (\ref{matching}) by the matrix $(i\tau-1/2\tau\mathcal{G}\tau)$
we obtain the matching conditions for the region outside the spherical shell
\begin{equation}\label{eq:matchext}
\left[I+\frac{1}{2}\tau\mathcal{G}\right]\phi^{\sigma}(R_+)=0\;.
\end{equation}
It is clear that in this case the above matching conditions lead to confinement as the
spinor field propagating in the interior of the shell and the one propagating in the exterior are completely independent. In this situation the potential models an \emph{impenetrable wall}.
\end{itemize}
Before we proceed with the analysis of the spectral zeta function we would like to provide a physical interpretation of the matching conditions (\ref{eq:matchint}) and (\ref{eq:matchext}) (c.f. \cite{stokes2015} and \cite{stokes2015a}). A $\delta$-shell potential is defined as \emph{impenetrable} if the spinors confined either in the interior or the exterior of the shell at $t=0$ remain confined in their respective regions for all $t\in\mathbb{R}^{+}$. In other words, the interior and exterior states are invariant under the time evolution operator \cite{dittrich1989dirac,arriza2015}. In the MIT bag conditions the matrix $\mathcal{G}$ in (\ref{matrixG}) reduces to $2\mathbb{I}$. In this 
case the impenetrability conditions (\ref{eq:matchint}) imply that the flux of the fermion current through the $\delta$-shell vanishes (see e.g. \cite{johnson1975}).
The MIT condition, then, simply generalizes to fermions the Dirichlet boundary condition since the latter is incompatible with the Dirac equation \cite{johnson1975}. 
The general matching conditions (\ref{eq:matchint}) and (\ref{eq:matchext}) describe, instead, all types of localized impenetrable potentials which keep interior and exterior 
fermions from traversing the $\delta$-shell. In particular, the general matching conditions (\ref{eq:matchint}) and (\ref{eq:matchext}) imply, just like in the MIT case, that
the flux of the fermion current through the shell vanishes. In fact, since the $\delta$-shell potential is static, in the rest frame of a point on the surface of the
$\delta$-shell, $n_i$ represents the ordinary space normal and $n_0=0$ \cite{johnson1975}. The matching conditions (\ref{eq:matchint}) applied to the total spinor $\varphi$ in (\ref{3}) (consisting of its radial and angular parts) are \cite{behrndt2019dirac} 
\begin{equation}
\left[i\sigma^j n_j+\frac{1}{2}\mathcal{G}\right]\varphi=0\;.
\end{equation}        
By using the fact that $(\sigma^j n_j)^{2}=\mathbb{I}$ we have
\begin{equation}\label{current1}
\left[\mathbb{I}+\frac{i}{2}\gamma^jn_j\tilde{\mathcal{G}}\right]\varphi=0\;,
\end{equation} 
where $\gamma^i=\beta \sigma^i$ and $\tilde{\mathcal{G}}=\beta\mathcal{G}$. By noticing that $\tilde{\mathcal{G}}^{\dagger}=\tilde{\mathcal{G}}$, the Dirac adjoint of the matching condition (\ref{current1}) has the form 
  \begin{equation}\label{current2}
\bar{\varphi}\left[\mathbb{I}-\frac{i}{2}\tilde{\mathcal{G}}\gamma^jn_j\right]=0\;,
\end{equation} 
with $\bar{\varphi}=\varphi^{\dagger}\beta$. By multiplying (\ref{current1}) by $\bar{\varphi}\tilde{\mathcal{G}}/2$ from the left and (\ref{current2}) by $\tilde{\mathcal{G}}\varphi/2$ from the right and by adding the resulting expressions we obtain
\begin{equation}\label{current3}
\frac{i}{2}n_j [\bar{\psi}\gamma^j \psi]=0\;,
\end{equation}
where we have defined $\psi=\tilde{\mathcal{G}}\varphi$. Let us, now, write explicitly the current that appears in (\ref{current3}). By noticing that 
$\tilde{\mathcal{G}}=g_s\mathbb{I}+g_v\beta$ we have
\begin{equation}\label{current4}
\bar{\psi}\gamma^j \psi=\bar{\varphi}[g_s^2\gamma^j+g_sg_v(\gamma^j\beta+\beta\gamma^j)-g^2_v\gamma^j]\varphi=(g_s^2-g_v^2)\bar{\varphi}\gamma^i\varphi\;.
\end{equation} 
By substituting this result in (\ref{current3}) and by recalling that in the confinement case $g_s^2-g_v^2=4$ we finally obtain
\begin{equation}\label{current5}
2i n_j \bar{\varphi}\gamma^j \varphi=0\;,
\end{equation} 
which implies that the flux of the fermion current through the $\delta$-shell vanishes. Obviously a similar interpretation is valid for the matching conditions (\ref{eq:matchext}).
In summary, the matching conditions (\ref{eq:matchint}) and (\ref{eq:matchext}) represent all possible constraints that can be imposed on a spinor field which 
will not allow any fermion current through the $\delta$-shell.

In this paper we will focus our attention to the \emph{confining case} since it is 
mathematically less involved and it contains, as a particular case, the MIT bag model. 
In order to analyze the Casimir energy of this system we utilize the spectral zeta function 
constructed from the eigenvalues of the equation (\ref{eq:eigenvalueeq}) endowed with 
the matching conditions (\ref{eq:matchint}) and (\ref{eq:matchext}). Since in the confining case the interior and exterior of the spherical shell are independent, we have two sets 
of eigenvalues that define an interior and an exterior spectral zeta function as follows:
\begin{equation}\label{zeta}
\zeta^{(\textrm{int}\backslash\textrm{ext})}(s)=\sum_{E^{(\textrm{int}\backslash\textrm{ext})}}E^{-2s}\;.
\end{equation} 
According to the general theory of spectral functions \cite{kirsten2001spectral}, (\ref{zeta}) is analytic in the semi-plane $\Re(s)>3/2$ and can be analytically extended to 
the left of the abscissa of convergence $\Re(s)=3/2$ to a meromorphic function with only isolated simple poles. Once the analytically continued expression for the spectral zeta function is obtained, one can compute the Casimir energy by following, for instance, \cite{kirsten2001spectral,bordag2009advances}.

The eigenvalues needed for the construction of the interior and exterior zeta functions 
can be obtained, implicitly, from the matching conditions (\ref{eq:matchint}) and (\ref{eq:matchext}), respectively. The matrix in the matching conditions (\ref{eq:matchint}) is singular since $\gp\gm+4=0$. This implies that the two equations it generates are equivalent. The same exact argument holds for (\ref{eq:matchext}). By utilizing a simple row reduction argument one can prove that the conditions (\ref{eq:matchint}) and (\ref{eq:matchext}) reduce to
\begin{equation}\label{eq:matchintsimp}
\left(
	\begin{array}{cc}
		1 & \pm \frac{i}{2} \gm \\
		0 & 0 \\
	\end{array}
	\right)\phi^{\sigma}(R_{\pm})=0\;,
\end{equation}
which, by setting $\gm=-2$, reduces to the MIT bag boundary conditions. The explicit form of the matching conditions (\ref{eq:matchintsimp}) allows us to prove the following two results.  
 
\begin{lemma}
The characteristic function for the spinor field associated with the interior region $r\in(0,R)$ is given by
\begin{equation}\label{eq:modegenint}
f_{\sigma}(p,R)=J_{j-\frac{\sigma-1}{2}}(p R)+\frac{\sigma \gm}{2}\sqrt{\frac{E_-}{E_+}} J_{j+\frac{\sigma+1}{2}}(p R)\;.
\end{equation}
The eigenvalues of the Dirac Hamiltonian in the interior region are the real zeroes of $f_{\sigma}(p,R)$.
\end{lemma} 
\begin{proof} In the interior of the spherical shell, the appropriate cylinder functions to choose are the Bessel functions of the first kind since they are finite at $r=0$. This implies that the interior eigenstates are given by
\begin{equation}\label{eq:eigenint}
\phi_\textrm{int}^{\sigma}(r)=\frac{1}{\sqrt{r}}\left(
	\begin{array}{c}
	i	 J_{j-\frac{\sigma-1}{2}}(p r)  \\
		-\sigma\sqrt{\frac{E_-}{E_+}} J_{j+\frac{\sigma+1}{2}}(p r) \\
	\end{array}
	\right)\;.
\end{equation} 
By imposing the interior matching conditions from (\ref{eq:matchintsimp}) we obtain the characteristic function (\ref{eq:modegenint}).
The zeroes of (\ref{eq:modegenint}) provide, then, the eigenvalues of (\ref{eq:eigenvalueeq}) in the interior of the spherical shell.
\end{proof}

The characteristic function associated with the exterior of the spherical shell is slightly more involved to obtain. In fact, we have  
\begin{lemma}
The characteristic function for the spinor field associated with the exterior region $r\in(R,\infty)$ is given by
\begin{equation}\label{modegenext1}
\mathcal{F}_{\sigma}(p,R,\bar{R})=\frac{f_{\sigma}^{(1)}(pR)g_{\sigma}^{(2)}(p\bar{R})-f_{\sigma}^{(2)}(pR)g_{\sigma}^{(1)}(p\bar{R})}{g^{(2)}_{\sigma}(p\bar{R})-g^{(1)}_{\sigma}(p\bar{R})}\;,
\end{equation}
where 
\begin{eqnarray}
f_{\sigma}^{(i)}(pR)&\!\!=\!\!&h^{(i)}_{j-\frac{\sigma-1}{2}}(p R)-\frac{\sigma \gm}{2}\sqrt{\frac{E_-}{E_+}}h^{(i)}_{j+\frac{\sigma+1}{2}}(p R)\;,\nonumber\\
g_{\sigma}^{(i)}(p\bar{R})&\!\!=\!\!&h^{(i)}_{j-\frac{\sigma-1}{2}}(p \bar{R})-\sigma\sqrt{\frac{E_-}{E_+}}h^{(i)}_{j+\frac{\sigma+1}{2}}(p \bar{R})\;,
\end{eqnarray}
for $i=\{1,2\}$. The eigenvalues of the Dirac Hamiltonian in the exterior region are the real zeroes of $\mathcal{F}_{\sigma}(p,R,\bar{R})$ when $\bar{R}\to\infty$.
\end{lemma}
\begin{proof}
For the exterior of the spherical shell, the appropriate cylinder functions to consider 
are combinations of Riccati-Hankel functions of first and second kind which represent outgoing and incoming spherical waves. The eigenstates are, therefore, written in the form
\begin{equation}\label{eq:eigenext}
\phi_\textrm{ext}^{\sigma}(r)=\frac{1}{\sqrt{r}}\left(
	\begin{array}{c}
	i	\left[ \mathcal{A}h^{(1)}_{j-\frac{\sigma-1}{2}}(p r)+\mathcal{B}h^{(2)}_{j-\frac{\sigma-1}{2}}(p r) \right] \\
		-\sigma\sqrt{\frac{E_-}{E_+}}\left[ \mathcal{A}h^{(1)}_{j+\frac{\sigma+1}{2}}(p r)+\mathcal{B}h^{(2)}_{j+\frac{\sigma+1}{2}}(p r) \right] \\
	\end{array}
	\right)\;.
\end{equation}  
We impose now the interior matching conditions from (\ref{eq:matchintsimp}) to obtain the 
relation
\begin{equation}\label{ext1}
\mathcal{A}h^{(1)}_{j-\frac{\sigma-1}{2}}(p R)+\mathcal{B}h^{(2)}_{j-\frac{\sigma-1}{2}}(p R)
-\frac{\sigma \gm}{2}\sqrt{\frac{E_-}{E_+}}\left[ \mathcal{A}h^{(1)}_{j+\frac{\sigma+1}{2}}(p R)+\mathcal{B}h^{(2)}_{j+\frac{\sigma+1}{2}}(p R) \right]=0\;.
\end{equation}
Since the region outside the spherical shell is unbounded, the spectrum of the Hamiltonian
in (\ref{eq:eigenvalueeq}) is continuous \cite{dittrich1989dirac}. In order to obtain a discrete spectrum and, consequently, be able to 
construct the spectral zeta function associated with the region outside the spherical shell, 
we enclose the entire system inside a large sphere of radius $r=\bar{R}$ \cite{bordag1996sphere,fucci2016sphere}. The auxiliary large sphere renders the outside region finite which allows us to obtain a characteristic function for the exterior eigenvalues once 
boundary conditions are imposed at $r=\bar{R}$. The specific boundary conditions to impose 
on the large sphere are irrelevant. In fact, once the spectral zeta function for the exterior region is constructed, one takes the limit $\bar{R}\to\infty$ which is independent of the boundary conditions imposed at $r=\bar{R}$ \cite{bordag1996sphere}. By imposing, for simplicity, MIT bag boundary conditions on $\phi_\textrm{ext}^{\sigma}(r)$ at $r=\bar{R}$
we obtain a second relation
\begin{equation}\label{ext2}
\mathcal{A}h^{(1)}_{j-\frac{\sigma-1}{2}}(p \bar{R})+\mathcal{B}h^{(2)}_{j-\frac{\sigma-1}{2}}(p \bar{R})
-\sigma \sqrt{\frac{E_-}{E_+}}\left[ \mathcal{A}h^{(1)}_{j+\frac{\sigma+1}{2}}(p \bar{R})+\mathcal{B}h^{(2)}_{j+\frac{\sigma+1}{2}}(p \bar{R}) \right]=0\;.
\end{equation}
The linear system consisting of (\ref{ext1}) and (\ref{ext2}) has a non-trivial solutions 
for the coefficients $\mathcal{A}$ and $\mathcal{B}$ if  
\begin{equation}\label{numerator}
f_{\sigma}^{(1)}(pR)g_{\sigma}^{(2)}(p\bar{R})-f_{\sigma}^{(2)}(pR)g_{\sigma}^{(1)}(p\bar{R})=0\;. 
\end{equation}
In the limit $\bar{R}\to\infty$, the relation (\ref{numerator}) does not yet provide implicitly the exterior eigenvalues of the $\delta$-shell system. This occurs because in the limit $\bar{R}\to\infty$ one obtains a continuous contribution to the spectrum that originates from the solutions of the free Dirac Hamiltonian \cite{kirsten2001spectral,bordag1996sphere}. To eliminate this spurious contribution, 
we impose MIT bag boundary conditions on the free solution of the Dirac Hamiltonian at $r=\bar{R}$ \cite{kirsten2001spectral} to obtain the relation 
\begin{equation}
g^{(2)}_{\sigma}(p\bar{R})-g^{(1)}_{\sigma}(p\bar{R})=0\;,
\end{equation} 
which generates the continuous spectrum as $\bar{R}\to\infty$. 
The discrete exterior spectrum is then obtained in the limit $\bar{R}\to\infty$ 
from the zeroes of the characteristic function in (\ref{modegenext1}).
\end{proof}
The results presented above can be used to obtain the expression of the interior and exterior spectral zeta function according to the following  

\begin{theorem}
The interior and exterior spectral zeta function associated with the Dirac Hamiltonian endowed with the $\delta$-shell potential are, respectively,
\begin{eqnarray}\label{zetaint}
\zeta^{(\textrm{int})}(s)=\frac{1}{2\pi i}\sum^{\infty}_{j=1/2,3/2,\cdots}(2j+1)\int_{\gamma^{(\textrm{int})}}(p^{2}+m^{2})^{-s}\frac{\partial}{\partial p}\ln\left[p^{-4j}\prod_{\sigma=\pm 1}f_{\sigma}(p,R)f^{(\textrm{ant})}_{\sigma}(p,R)\right] \dpp \;,
\end{eqnarray}
and 
\begin{eqnarray}\label{zetaext}
\zeta^{(\textrm{ext})}(s)&=&\lim_{\bar{R}\to\infty}\frac{1}{2\pi i}\sum^{\infty}_{j=1/2,3/2,\cdots}(2j+1)\int_{\gamma^{(\textrm{ext})}}(p^{2}+m^{2})^{-s}\nonumber\\
&\times& \frac{\partial}{\partial p}\ln\left[p^{4(j+1)}\prod_{\sigma=\pm 1}\mathcal{F}_{\sigma}(p,R,\bar{R})\mathcal{F}^{(\textrm{ant})}_{\sigma}(p,R,\bar{R})\right] \dpp \;,
\end{eqnarray}
where $\gamma^{(\textrm{int})}$ and $\gamma^{(\textrm{ext})}$ represent contours that enclose, in the counterclockwise direction, all the zeroes of the total characteristic function in square parentheses in (\ref{zetaint}) and in (\ref{zetaext}), respectively. The integral representations (\ref{zetaint}) and (\ref{zetaext}) are valid, by construction, in the region $\Re(s)>3/2$.
\end{theorem}
\begin{proof}
As we have shown in the previous lemmas, the characteristic functions found in (\ref{eq:modegenint}) and (\ref{modegenext}) 
provide the eigenvalues of the Dirac Hamiltonian for the spinor field. However, in 
order to construct the appropriate spectral zeta function used to compute the Casimir energy of the system, we need to include 
the eigenvalues associated with the anti-spinor field as well. This is not a very 
difficult task since we can obtain the relevant expressions for the anti-spinor field 
by simply performing the replacement $\sigma\to-\sigma$, $E\to -E$ and $g_v\to-g_v$ in the 
corresponding expressions for the spinor field as proved in \cite{grant2007relativistic}.
This implies that the interior and exterior characteristic functions for anti-spinors, namely $f^{(\textrm{ant})}_{\sigma}(p,R)$ and $\mathcal{F}^{(\textrm{ant})}_{\sigma}(p,R,\bar{R})$ are obtained from (\ref{eq:modegenint}) and from (\ref{modegenext1}), respectively, by exploiting the replacement $\sigma\to-\sigma$ and $E\to -E$. 
The \emph{total} interior and exterior characteristic functions are obtained as the product of the one for spinors and the one for anti-spinors in the interior and exterior regions, respectively. 
By using the argument principle, as outlined, for instance, in \cite{kirsten2001spectral}, we obtain the interior and exterior spectra zeta functions (\ref{zetaint}) and (\ref{zetaext}) in terms of the \emph{total} characteristic functions. The prefactors $p^{-4j}$ in (\ref{zetaint}) and $p^{4(j+1)}$ in (\ref{zetaext}) are needed in order to avoid contributions coming from the origin in the process of contour deformation (see e.g. \cite{kirsten2001spectral}). They are obtained by analyzing the small-$p$ behavior of the \emph{total} characteristic functions.   
\end{proof}

We would like to point out that since MIT boundary conditions are obtained by setting 
$g_s=2$ and $g_{v}=0$, the characteristic functions satisfy the simple relations 
\begin{eqnarray*}
f^{(\textrm{ant})}_{\sigma=1}(p,R)f^{(\textrm{ant})}_{\sigma=-1}(p,R)&=&f_{\sigma=1}(p,R)f_{\sigma=-1}(p,R),\\[1ex]
f^{(\textrm{ant})}_{\sigma=1}(p,R)f_{\sigma=1}(p,R)&=&f^{(\textrm{ant})}_{\sigma=-1}(p,R)f_{\sigma=-1}(p,R).
\end{eqnarray*}
This implies that in order to include the anti-spinor contribution to the zeta function in the MIT case it is sufficient to simply multiply the spinor characteristic function by a factor of 2 \cite{elizalde1998casimir,bordag2009advances}.

\section{Analytic continuation of the spectral zeta function}

As we will point point out in the next Section, in order to evaluate the Casimir energy of the system we need to analytically continue the representations (\ref{zetaint}) and (\ref{zetaext}) to a neighborhood of the point $s=-1/2$. The process of analytic continuation of the integral representations (\ref{zetaint}) and (\ref{zetaext}) is rather standard (see e.g. \cite{kirsten2001spectral}) and consists, essentially, of two steps. 
The first is a deformation of the integration contour to the imaginary axis, the result of which is contained in the following 
\begin{lemma}
The interior and exterior spectral zeta function possess the integral representations
\begin{eqnarray}\label{zetaint1}
\lefteqn{\zeta^{(\textrm{int})}(s)=\frac{\sin(\pi s)}{\pi}\sum^{\infty}_{j=1/2,3/2,\cdots}(2j+1)\sum_{\sigma=\pm 1}\int_{m}^{\infty}(p^{2}-m^{2})^{-s}\frac{\partial}{\partial p}}\\
 &&\times\ln\left[p^{-2j}\left(I_{j-\frac{\sigma-1}{2}}^{2}(pR)+\frac{\gm^{2}}{4}I_{j+\frac{\sigma+1}{2}}^{2}(pR)-\frac{m \gm}{p}I_{j-\frac{\sigma-1}{2}}(pR)I_{j+\frac{\sigma+1}{2}}(pR)\right)\right] \dpp \;,\nonumber
\end{eqnarray}
and
\begin{eqnarray}\label{zetaext3}
\lefteqn{\zeta^{(\textrm{ext})}(s)=\frac{\sin(\pi s)}{\pi}\sum^{\infty}_{j=1/2,3/2,\cdots}(2j+1)\sum_{\sigma=\pm 1}\int_{m}^{\infty}(p^{2}-m^{2})^{-s}\frac{\partial}{\partial p}}\\
 &&\times\ln\left[p^{2j+2}\left(K_{j-\frac{\sigma-1}{2}}^{2}(pR)+\frac{\gm^{2}}{4}K_{j+\frac{\sigma+1}{2}}^{2}(pR)-\frac{m \gm}{p}K_{j-\frac{\sigma-1}{2}}(pR)K_{j+\frac{\sigma+1}{2}}(pR)\right)\right] \dpp \;,\nonumber
\end{eqnarray}
which are valid in the strip $1/2<\Re(s)<1$.
\end{lemma}
\begin{proof}
For the interior region, by noticing that 
\begin{eqnarray}\label{eq:mode_gen}
\lefteqn{f_{\sigma}\left(e^{\pm \frac{i\pi}{2}}p, R\right)f^{(\textrm{ant})}_{-\sigma}\left(e^{\pm \frac{i\pi}{2}}p, R\right)}\\
&&=\sigma e^{\pm i\pi}\left[I_{j-\frac{\sigma-1}{2}}^{2}(pR)+\frac{\gm^{2}}{4}I_{j+\frac{\sigma+1}{2}}^{2}(pR)-\frac{m \gm}{p}I_{j-\frac{\sigma-1}{2}}(pR)I_{j+\frac{\sigma+1}{2}}(pR)\right]\;,\nonumber
\end{eqnarray}
where $I_{\nu}(x)$ represents the modified Bessel function of the first kind, a deformation to the imaginary axis of the integration contour $\gamma^{(\textrm{int})}$ in (\ref{zetaint}) leads to the  integral representation (\ref{zetaint1}).

For the spectral zeta function associated with the outside of the spherical shell, a deformation of the integration contour $\gamma^{(\textrm{int})}$ to the imaginary axis leads to the expression
\begin{eqnarray}\label{zetaext1}
\lefteqn{\zeta^{(\textrm{ext})}(s,\bar{R})=\sum_{\mu=\pm 1}\frac{\mu e^{\mu i \pi s}}{2\pi i}\sum^{\infty}_{j=1/2,3/2,\cdots}(2j+1)\int_{m}^{\infty}(p^{2}-m^{2})^{-s}\frac{\partial}{\partial p}}\\
&& \times \ln\left[\left(e^{-\frac{\mu i\pi}{2}}p\right)^{4j+2}\prod_{\sigma=\pm 1}\mathcal{F}_{\sigma}\left(e^{-\frac{\mu i\pi}{2}}p,R,\bar{R}\right)\mathcal{F}^{(\textrm{ant})}_{\sigma}\left(e^{-\frac{\mu i\pi}{2}}p,R,\bar{R}\right)\right] \dpp \;.\nonumber
\end{eqnarray}
At this stage we can perform the limit $\bar{R}\to\infty$ as required in (\ref{zetaext}).
By using the following asymptotic behavior of the Riccati-Hankel functions (\cite{olver2010nist}, Section 10.17) 
\begin{eqnarray}
h^{(\alpha)}_{\nu}\left(e^{\frac{\mu i\pi}{2}}x\right)&\!\!=\!\!&e^{(-1)^{\alpha}\mu x}\left[1+O\left(\frac{1}{x}\right)\right]\;,
\end{eqnarray}  
for $\alpha=\{1,2\}$ and $x\in\mathbb{R}$, one can show that
\begin{eqnarray}
\lim_{\bar{R}\to\infty}\mathcal{F}_{\sigma}\left(e^{\frac{ i\pi}{2}}p,R,\bar{R}\right)&\!\!=\!\!&f_{\sigma}^{(1)}\left(e^{\frac{ i\pi}{2}}pR\right)\;,\nonumber\\
\lim_{\bar{R}\to\infty}\mathcal{F}_{\sigma}\left(e^{-\frac{ i\pi}{2}}p,R,\bar{R}\right)&\!\!=\!\!&f_{\sigma}^{(2)}\left(e^{-\frac{i\pi}{2}}pR\right)\;,
\end{eqnarray}
and similar limits hold for $\mathcal{F}^{(\textrm{ant})}_{\sigma}(e^{-\frac{\mu i\pi}{2}}p,R,\bar{R})$. 
The representation (\ref{zetaext1}) together with the above limits allows us to write
\begin{eqnarray}\label{zetaext2}
\lefteqn{\zeta^{(\textrm{ext})}(s)=-\frac{e^{-i \pi s}}{2\pi i}\sum^{\infty}_{j=1/2,3/2,\cdots}(2j+1)\int_{m}^{\infty}(p^{2}-m^{2})^{-s}\frac{\partial}{\partial p}}\\
& \times & \ln\left[\left(e^{\frac{ i\pi}{2}}p\right)^{4j+2}\prod_{\sigma=\pm 1}f^{(1)}_{\sigma}\left(e^{\frac{i\pi}{2}}p,R\right)f^{(1),(\textrm{ant})}_{\sigma}\left(e^{\frac{i\pi}{2}}p,R\right)\right] \dpp \nonumber\\
&+&\frac{e^{i \pi s}}{2\pi i}\sum^{\infty}_{j=1/2,3/2,\cdots}(2j+1)\int_{m}^{\infty}(p^{2}-m^{2})^{-s}\frac{\partial}{\partial p}\nonumber\\
& \times & \ln\left[\left(e^{-\frac{i\pi}{2}}p\right)^{4j+2}\prod_{\sigma=\pm 1}f^{(2)}_{\sigma}\left(e^{-\frac{i\pi}{2}}p,R\right)f^{(2),(\textrm{ant})}_{\sigma}\left(e^{-\frac{i\pi}{2}}p,R\right)\right]\dpp\;.\nonumber
\end{eqnarray}
By noticing that 
\begin{eqnarray}\label{2022}
f_{\sigma}^{(i)}\left(e^{\pm\frac{i\pi}{2}}p,R\right)f_{-\sigma}^{(i),(\textrm{ant})}\left(e^{\pm\frac{i\pi}{2}}p,R\right)&=&\left[h^{(i)}_{j-\frac{\sigma-1}{2}}\left(e^{\pm\frac{i\pi}{2}}pR\right)\right]^{2}-\frac{\gm^{2}}{4}\left[h^{(i)}_{j+\frac{\sigma+1}{2}}\left(e^{\pm\frac{i\pi}{2}}pR\right)\right]^{2}\nonumber\\
&&\pm\sigma\frac{m \gm}{i p}h^{(i)}_{j-\frac{\sigma-1}{2}}\left(e^{\pm\frac{i\pi}{2}}pR\right)h^{(i)}_{j+\frac{\sigma+1}{2}}\left(e^{\pm\frac{i\pi}{2}}pR\right)\;,
\end{eqnarray}
and by using the relations \cite{olver2010nist}
\begin{eqnarray}
h^{(1)}_{\nu}\left(e^{\frac{i\pi}{2}}pR\right)=\frac{2}{i\pi}e^{-\frac{i\pi\nu}{2}}K_{\nu}(pR)\;,\nonumber\\
h^{(2)}_{\nu}\left(e^{-\frac{i\pi}{2}}pR\right)=-\frac{2}{i\pi}e^{\frac{i\pi\nu}{2}}K_{\nu}(pR)\;,
\end{eqnarray}
where $K_{\nu}(z)$ denotes the modified Bessel function of the second kind, one can obtain
(\ref{zetaext3}) from (\ref{zetaext2}).
\end{proof}
It is instructive to notice that in the particular case of the MIT bag conditions, namely when $\gm=-2$, the contributions to the integral representation (\ref{zetaint1}) coming from the two values of $\sigma$ are identical and therefore it reduces to the same integral representation found in \cite{elizalde1998casimir}. 
A similar occurrence happens in the exterior case; in fact, when $\gm=-2$ the expression (\ref{zetaext3}) reduces to the result obtained in \cite{elizalde1998casimir} for the MIT bag conditions. In addition, it is interesting to notice that the spectral zeta function for the exterior region can be obtained from the one in the interior region by simply performing the substitution $I_{\nu}(z)\to K_{\nu}(z)$, a fact that was also mentioned in \cite{elizalde1998casimir}.

The second step in the process of analytic continuation consists in the subtraction, and then addition, of a suitable number of terms of the uniform asymptotic expansion of the derivative
appearing in the integral representations (\ref{zetaint1}) and (\ref{zetaext3}). This process leads to the 
\begin{theorem}
Let $N\in\mathbb{N}_{+}$. The interior and exterior spectral zeta functions are expressed, in the semi-plane $\Re(s)>(3-N)/2$, as  
\begin{equation}\label{zetacontinued}
\zeta^{(\textrm{int})/(\textrm{ext})}(s)=Z^{(\textrm{int})/(\textrm{ext})}(s)
+\sum_{i=-1}^{N}A^{(\textrm{int})/(\textrm{ext})}_{i}(s)\;.
\end{equation}
where $Z^{(\textrm{int})/(\textrm{ext})}(s)$ are analytic functions for $\Re(s)>(3-N)/2$ given in (\ref{zedint}) and (\ref{zedext}) and
$A^{(\textrm{int})/(\textrm{ext})}_{i}(s)$ are meromorphic functions of $s\in\mathbb{C}$ with isolated simple poles provided in (\ref{aminus1int})-(\ref{aiint}) and (\ref{aminus1ext})-(\ref{aiext}).  
\end{theorem} 
\begin{proof}
In order to evaluate the asymptotic expansions, we need to use the relations \cite{olver2010nist}
(Chapter 10, Section 29)
\begin{equation}\label{trans}
	\begin{aligned}
K_{\nu+1}(z)&=&\frac{\nu}{z}K_{\nu}(z)-K'_{\nu}(z)\;,\\[1ex]
I_{\nu+1}(z)&=&-\frac{\nu}{z}I_{\nu}(z)+I'_{\nu}(z)\;,
	\end{aligned}
\end{equation} 
so that the modified Bessel functions with a $j+1$ index appearing in (\ref{zetaint1}) and (\ref{zetaext3}) can be rewritten in terms of quantities with only a $j$ index. More explicitly, by using (\ref{trans}) in (\ref{zetaint1}) and (\ref{zetaext3}) one obtains, after the change of variables $pR\to z j$, the representations
\begin{eqnarray}\label{zetaintfinal}
\zeta^{(\textrm{int})/(\textrm{ext})}(s)=\frac{\sin(\pi s)}{\pi}\sum^{\infty}_{j=1/2,3/2,\cdots}(2j+1)\sum_{\sigma=\pm 1}\int_{\frac{m R}{j}}^{\infty}\left[\left(\frac{zj}{R}\right)^{2}-m^{2}\right]^{-s}\frac{\partial}{\partial z}\ln\left[ \mathcal{G}^{(\textrm{int})/(\textrm{ext})}_{j,\sigma}(z)\right] \dz\;,
\end{eqnarray} 
where we have defined
\begin{eqnarray}\label{modegenint}
\mathcal{G}^{(\textrm{int})}_{j,\sigma}(z)&=&z^{-2j}
\Bigg\{I_{j}^{2}(zj)
\left[
\left(\frac{\gm^{2}}{4}+\frac{1}{z^{2}}\right)
\sigma_{-}
+\left(1+\frac{\gm^{2}}{4z^{2}}\right)
\sigma_{+}
+\frac{mR \gm }{z^{2}j}
\right]\nonumber\\
&+&(I'_{j})^{2}(zj)\left[
\sigma_{-}+
\frac{\gm^{2}}{4}\sigma_{+}
\right]-\frac{2 }{z }I_{j}(zj)I'_{j}(zj)\left[
\sigma_{-}
+\frac{\gm^{2}}{4}
\sigma_{+}
+\frac{mR \gm}{2j}
\right] 
\Bigg\}\;,\nonumber\\
\end{eqnarray}
and
\begin{eqnarray}\label{modegenext}
\mathcal{G}^{(\textrm{ext})}_{j,\sigma}(z)&=&z^{2j+2}
\Bigg\{K_{j}^{2}(zj)
\left[
\left(\frac{\gm^{2}}{4}+\frac{1}{z^{2}}\right)
\sigma_{-}
+\left(1+\frac{\gm^{2}}{4z^{2}}\right)
\sigma_{+}
-\frac{mR \gm }{z^{2}j}
\right]\nonumber\\
&+&(K'_{j})^{2}(zj)\left[
\sigma_{-}+
\frac{\gm^{2}}{4}\sigma_{+}
\right]-\frac{2 }{z }K_{j}(zj)K'_{j}(zj)\left[
\sigma_{-}
+\frac{\gm^{2}}{4}
\sigma_{+}
-\frac{mR \gm}{2j}
\right] 
\Bigg\}\;,\nonumber\\
\end{eqnarray}
with 
\begin{equation}
\sigma_{\pm}=\frac{1\pm\sigma}{2}\;.
\end{equation}
By subtracting and then adding the first $N$ terms of the uniform asymptotic expansion of the characteristic functions in (\ref{modegenint}) and (\ref{modegenext}), derived in Appendix \ref{app}, we obtain the expression (\ref{zetacontinued}) for the interior and exterior spectral zeta functions.
The functions $Z^{(\textrm{int})/(\textrm{ext})}(s)$ read 
\begin{eqnarray}\label{zedint}
Z^{(\textrm{int})}(s)&=&\frac{\sin(\pi s)}{\pi}
\sum^{\infty}_{j=1/2,3/2,\cdots}(2j+1)\sum_{\sigma=\pm 1}\int_{\frac{m R}{j}}^{\infty}\left[\left(\frac{zj}{R}\right)^{2}-m^{2}\right]^{-s}\nonumber\\
&\times&\frac{\partial}{\partial z}\Bigg\{\ln\mathcal{G}^{(\textrm{int})}_{j,\sigma}(z)-
\ln\left[\frac{e^{2j\eta}(1+z^2)^{1/2}(1-t)}{\pi j z^{2j+2}}\Omega^{(\textrm{int})}_{0,\sigma}(t)\right]\nonumber\\
&-&\ln\left[1-\sum_{k=1}^{\infty}\frac{\Omega^{(\textrm{int})}_{k,\sigma}(t)}{\Omega^{(\textrm{int})}_{0,\sigma}(t)}j^{-k}\right]
\Bigg\}\dz
\end{eqnarray}
and
\begin{eqnarray}\label{zedext}
Z^{(\textrm{ext})}(s)&=&\frac{\sin(\pi s)}{\pi}
\sum^{\infty}_{j=1/2,3/2,\cdots}(2j+1)\sum_{\sigma=\pm 1}\int_{\frac{m R}{j}}^{\infty}\left[\left(\frac{zj}{R}\right)^{2}-m^{2}\right]^{-s}\nonumber\\
&\times&\frac{\partial}{\partial z}\Bigg\{\ln\mathcal{G}^{(\textrm{ext})}_{j,\sigma}(z)-
\ln\left[\frac{\pi e^{-2j\eta}(1+z^2)^{1/2}(1+t)}{j z^{-2j}}\Omega^{(\textrm{ext})}_{0,\sigma}(t)\right]\nonumber\\
&-&\ln\left[1-\sum_{k=1}^{\infty}\frac{\Omega^{(\textrm{ext})}_{k,\sigma}(t)}{\Omega^{(\textrm{ext})}_{0,\sigma}(t)}j^{-k}\right]
\Bigg\}\dz\;,
\end{eqnarray}
which are, by construction, analytic functions in the semi-plane $\Re(s)>(3-N)/2$. The remaining functions, $A^{(\textrm{int})/(\textrm{ext})}_{i}(s)$, are meromorphic in the complex plane and contain all the information about the simple poles of the spectral zeta function. The expressions for the interior functions are
\begin{eqnarray}\label{aminus1int}
A^{(\textrm{int})}_{-1}(s)&\!\!=\!\!&4\frac{\sin(\pi s)}{\pi}
\sum^{\infty}_{j=1/2,3/2,\cdots}j(2j+1)\int_{\frac{m R}{j}}^{\infty}\left[\left(\frac{zj}{R}\right)^{2}-m^{2}\right]^{-s}\left(\frac{\sqrt{1+z^{2}}-1}{z}\right)\dz\;,\\[1ex]
\label{a0int}
A^{(\textrm{int})}_{0}(s)&=&a_{0}(s)+b_{0}(s)\;,
\end{eqnarray}
where we have defined
\begin{equation}\label{a0intparts}
	\begin{aligned}
a_{0}(s)&=&-2\frac{\sin(\pi s)}{\pi}
\sum^{\infty}_{j=1/2,3/2,\cdots}(2j+1)\int_{\frac{m R}{j}}^{\infty}\left[\left(\frac{zj}{R}\right)^{2}-m^{2}\right]^{-s}\frac{\partial}{\partial z}\ln\left[\sqrt{1+z^{2}}+1\right] \dz\;,\\
b_{0}(s)&=&\frac{\sin(\pi s)}{\pi}
\sum^{\infty}_{j=1/2,3/2,\cdots}(2j+1)\sum_{\sigma=\pm 1}\int_{\frac{m R}{j}}^{\infty}\left[\left(\frac{zj}{R}\right)^{2}-m^{2}\right]^{-s}\frac{\partial}{\partial z}\ln\left[\Omega^{(\textrm{int})}_{0,\sigma}(t)\right] \dz\;.
	\end{aligned}
\end{equation}
For $i\geq 1$ we obtain, instead,
\begin{eqnarray}\label{aiint}
A^{(\textrm{int})}_{i}(s)=\frac{\sin(\pi s)}{\pi}
\sum^{\infty}_{j=1/2,3/2,\cdots}j^{-i}(2j+1)
\sum_{\sigma=\pm 1}\int_{\frac{m R}{j}}^{\infty}\left[\left(\frac{zj}{R}\right)^{2}-m^{2}\right]^{-s}\frac{\partial}{\partial z}\left[\mathcal{D}^{(\textrm{int})}_{i,\sigma}(t)\right] \dz\;,
\end{eqnarray}
where the functions $\mathcal{D}^{(\textrm{int})}_{i,\sigma}(t)$ can be determined by 
the cumulant expansion
\begin{eqnarray}\label{cumulint}
\ln\left[1-\sum_{k=1}^{\infty}\frac{\Omega^{(\textrm{int})}_{k,\sigma}(t)}{\Omega^{(\textrm{int})}_{0,\sigma}(t)}j^{-k}\right]\simeq \sum_{k=1}^{\infty}\frac{\mathcal{D}^{(\textrm{int})}_{k,\sigma}(t)}{j^{k}}\;.
\end{eqnarray}
For the exterior functions one finds, instead, 
\begin{eqnarray}\label{aminus1ext}
A^{(\textrm{ext})}_{-1}(s)&\!\!=\!\!&-A^{(\textrm{int})}_{-1}(s)\;, \\[0.5ex]
\label{a0ext}
A^{(\textrm{ext})}_{0}(s)&\!\!=\!\!&-a_{0}(s)+b_{0}(s)\;,
\end{eqnarray}
where $b_{0}(s)$ appears in the exterior case as well because of the relation (\ref{omegazeroext}) and the fact that $b_{0}(s)$ in (\ref{a0intparts}) contains the sum of both spin projections.
When $i\geq 1$, we get
\begin{eqnarray}\label{aiext}
A^{(\textrm{ext})}_{i}(s)=\frac{\sin(\pi s)}{\pi}
\sum^{\infty}_{j=1/2,3/2,\cdots}j^{-i}(2j+1)
\sum_{\sigma=\pm 1}\int_{\frac{m R}{j}}^{\infty}\left[\left(\frac{zj}{R}\right)^{2}-m^{2}\right]^{-s}\frac{\partial}{\partial z}\left[\mathcal{D}^{(\textrm{ext})}_{i,\sigma}(t)\right] \dz\;,
\end{eqnarray} 
where, similarly to the interior case, the functions $\mathcal{D}^{(\textrm{ext})}_{i,\sigma}(t)$ are found through the relation
\begin{eqnarray}\label{cumulext}
\ln\left[1-\sum_{k=1}^{\infty}\frac{\Omega^{(\textrm{ext})}_{k,\sigma}(t)}{\Omega^{(\textrm{ext})}_{0,\sigma}(t)}j^{-k}\right]\simeq \sum_{k=1}^{\infty}\frac{\mathcal{D}^{(\textrm{ext})}_{k,\sigma}(t)}{j^{k}}\;.
\end{eqnarray}
\end{proof}
Unfortunately, there is no closed-form expression for the integrals (\ref{aminus1int})-(\ref{aiint}) and (\ref{aminus1ext})-(\ref{aiext}) which is valid for all values of the mass $m$. We can, however, find a small-$m$ expansion. 
\begin{lemma}
The small-$m$ asymptotic expansion of the meromorphic terms in the analytic continuation of the interior and exterior spectral zeta function (\ref{zetacontinued})
is of the form
\begin{equation}
A^{(\textrm{int})/(\textrm{ext})}_{i}(s)=\sum_{k=0}^{\infty}\mathscr{A}^{(\textrm{int})/(\textrm{ext})}_{i,k}(s,R)(mR)^{2k}\;,
\end{equation}
where the coefficient functions $\mathscr{A}^{(\textrm{int})/(\textrm{ext})}_{i,k}(s,R)$ are provided in (\ref{a-1final}), (\ref{a0final}), (\ref{b0final}), (\ref{aifinal}), and (\ref{aifinalext}).    
\end{lemma}
\begin{proof}
Since the terms 
$A^{(\textrm{int})}_{-1}(s)$ and $a_{0}(s)$ also appear in the analytic continuation 
of the spectral zeta function for the MIT bag case, their small-$m$ expansion can be found in  \cite{elizalde1998casimir} and are obtained by following the procedure described in details in Section 3.1 of \cite{kirsten2001spectral}.
One, therefore, has 
\begin{eqnarray}\label{a-1final}
A^{(\textrm{int})}_{-1}(s)=\frac{R^{2s}}{\sqrt{\pi}\Gamma(s)}\sum_{k=0}^{\infty}\frac{(-1)^{k}(mR)^{2k}}{k!}\frac{\Gamma(k+s-1/2)}{k+s}\left[2\zeta_{H}\left(2s+2k-2;\frac{1}{2}\right)+\zeta_{H}\left(2s+2k-1;\frac{1}{2}\right)\right]\;,
\end{eqnarray}
and 
\begin{eqnarray}\label{a0final}
a_{0}(s)=-\frac{R^{2s}}{\sqrt{\pi}\Gamma(s)}\sum_{k=0}^{\infty}\frac{(-1)^{k}(mR)^{2k}}{k!}\frac{\Gamma(k+s+1/2)}{k+s}\left[2\zeta_{H}\left(2s+2k-1;\frac{1}{2}\right)+\zeta_{H}\left(2s+2k;\frac{1}{2}\right)\right]\;.
\end{eqnarray}
In order to obtain the small-$m$ expansion of the remaining functions, we need to use the results outlined in the Appendices.
By rewriting the integral of $b_{0}(s)$ in (\ref{a0intparts}) in terms of the variable $t$ and by then performing the change of variables $t=t_{j}u$, with $t_{j}$ given in 
(\ref{massint}), one obtains
\begin{eqnarray}\label{intbzero}
b_{0}(s)=-\frac{\sin(\pi s)}{\pi}R^{2s}
\sum^{\infty}_{j=1/2,3/2,\cdots}j^{-2s}(2j+1)t_{j}^{2s}\sum_{\sigma=\pm 1}\int_{0}^{1}u^{2s}\left(1-u^{2}\right)^{-s}\frac{\partial}{\partial u}\ln\left[\Omega^{(\textrm{int})}_{0,\sigma}(t_{j}u)\right] \du\;.
\end{eqnarray}    
By recalling the form of $\Omega^{(\textrm{int}}_{0,\sigma}(z)$ in (\ref{omegazeroint}), we find
\begin{equation}\label{logomegazero}
\frac{\partial}{\partial u}\ln\left[\Omega^{(\textrm{int}/\textrm{ext})}_{0,\sigma}(t_{j}u)\right] =\frac{\left(4-\gm^{2}\right)t_{j}}{\sigma\left(4+\gm^{2}\right)+\left(4-\gm^{2}\right)t_{j} u}\;,
\end{equation}
and the integral in (\ref{intbzero}) becomes
\begin{eqnarray}\label{intbzero1}
b_{0}(s)=-\frac{\sin(\pi s)}{\pi}R^{2s}
\sum^{\infty}_{j=1/2,3/2,\cdots}j^{-2s}(2j+1)\sum_{\sigma=\pm 1}\beta_{\sigma}t_{j}^{2s+1}\int_{0}^{1}u^{2s}\left(1-u^{2}\right)^{-s}
\left(1+\beta_{\sigma}t_{j}u\right)^{-1} \du\;.
\end{eqnarray}
where
\begin{equation}
\beta_{\sigma}=\sigma\frac{4-\gm^{2}}{4+\gm^{2}}
\end{equation}
It is not very difficult to realize that the integral in (\ref{intbzero1}) is of the form
(\ref{integral}) with $n=1$ and $\alpha=0$. By using the result (\ref{exp3}) we get
\begin{eqnarray}
b_{0}(s)=-\frac{\sin(\pi s)}{\pi}R^{2s}
\sum^{\infty}_{j=1/2,3/2,\cdots}j^{-2s}(2j+1)\sum_{\sigma=\pm 1}\beta_{\sigma}\sum_{p=0}^{\infty}\Lambda_{p}(s,0,1,\beta_{\sigma})
\left(\frac{mR}{j}\right)^{2p}\;.
\end{eqnarray}
The sum over the parameter $j$ can be computed in terms of the Hurwitz zeta function to obtain
\begin{eqnarray}\label{b0final}
b_{0}(s)=-\frac{\sin(\pi s)}{\pi}R^{2s}\sum_{\sigma=\pm 1}\beta_{\sigma}\sum_{p=0}^{\infty}\Lambda_{p}(s,0,1,\beta_{\sigma})
(mR)^{2p}\left[\zeta_{H}\left(2s+2p-1,\frac{1}{2}\right)+\zeta_{H}\left(2s+2p,\frac{1}{2}\right)\right]\;.
\end{eqnarray}
The small-$m$ expansion of the functions $A^{(\textrm{int})}_{i}(s)$ can be found by following the computation outlined for $b_{0}(s)$. From Appendix \ref{appb} one 
finds that the general form of the functions $\mathcal{D}^{(\textrm{int})}_{k,\sigma}(t)$ is
\begin{equation}\label{generalD}
\mathcal{D}^{(\textrm{int})}_{k,\sigma}(t)=\frac{2^{2k}(4+\gm^{2})^{-k}t^{k}}{(1+\beta_{\sigma}t)^{k}}\sum_{l=0}^{3k}c^{(\textrm{int})}_{k,l}(mR,\sigma)t^{l}\;,
\end{equation}
where the coefficients $c^{(\textrm{int})}_{k,l}(mR,\sigma)$ can be read off the expressions for $\mathcal{D}^{(\textrm{int})}_{k,\sigma}(t)$ in Appendix \ref{app}, and can be written as
\begin{equation}\label{cees}
c^{(\textrm{int})}_{k,l}(mR,\sigma)=\sum_{q=0}^{k}d^{(\textrm{int})}_{q,\sigma}(k,l)(mR)^{q}\;.
\end{equation}
By re-expressing the integral for 
$A^{(\textrm{int})}_{i}(s)$ in (\ref{aiint}) in terms $t$, and by subsequently changing variable $t\to t_{j}u$, we obtain 
\begin{eqnarray}
A^{(\textrm{int})}_{i}(s)&\!\!=\!\!&\frac{\sin(\pi s)}{2^{2i}\pi}R^{2s}(4+\gm^{2})^{-i}
\sum^{\infty}_{j=1/2,3/2,\cdots}j^{-2s-i}(2j+1)\\
&&\times\sum_{\sigma=\pm 1}\sum_{l=0}^{3i}c^{(\textrm{int})}_{i,l}(mR,\sigma)\Big[i\beta_{\sigma}I_{i+1}(s,i+l)-(l+i)I_{i}(s,i+l-1)\Big]\;.\nonumber
\end{eqnarray}
The relation (\ref{exp3}) allows us to get the following expression
\begin{eqnarray}
A^{(\textrm{int})}_{i}(s)&\!\!=\!\!&\frac{\sin(\pi s)}{2^{2i}\pi}R^{2s}(4+\gm^{2})^{-i}
\sum^{\infty}_{j=1/2,3/2,\cdots}j^{-2s-i}(2j+1)\nonumber\\
&&\times\sum_{\sigma=\pm 1}\sum_{l=0}^{3i}c^{(\textrm{int})}_{i,l}(mR,\sigma)\sum_{p=0}^{\infty}\left(\frac{mR}{j}\right)^{2p}
\Big[i\beta_{\sigma}\Lambda_{p}(s,i+l,i+1,\beta_{\sigma})\nonumber\\
&&-(l+i)\Lambda_{p}(s,i+l-1,i,\beta_{\sigma})\Big]\;.
\end{eqnarray}
The sum over $j$ can be performed and expressed in terms of the Hurwitz zeta function to obtain
\begin{eqnarray}
A^{(\textrm{int})}_{i}(s)&\!\!=\!\!&\frac{\sin(\pi s)}{2^{2i}\pi}R^{2s}(4+\gm^{2})^{-i}
\sum_{n=0}^{\infty}(mR)^{2n}\tau^{(\textrm{int})}_{n,i}(s,mR)\nonumber\\
&&\times \left[2\zeta_{H}\left(2s+i+2n-1,\frac{1}{2}\right)+\zeta_{H}\left(2s+i+2n,\frac{1}{2}\right)\right]\;,
\end{eqnarray}
with 
\begin{eqnarray}
\tau^{(\textrm{int})}_{n,i}(s,mR)=\sum_{\sigma=\pm 1}\sum_{l=0}^{3i}c^{(\textrm{int})}_{i,l}(mR,\sigma)\Big[i\beta_{\sigma}\Lambda_{n}(s,i+l,i+1,\beta_{\sigma})-(l+i)\Lambda_{n}(s,i+l-1,i,\beta_{\sigma})\Big]\;.
\end{eqnarray}
By recalling from (\ref{cees}) that $c^{(\textrm{int})}_{i,l}(mR,\sigma)$ is a polynomial in $mR$, we can finally write the small-$m$ expansion
\begin{eqnarray}\label{aifinal}
A^{(\textrm{int})}_{i}(s)&\!\!=\!\!&\frac{\sin(\pi s)}{2^{2i}\pi}R^{2s}(4+\gm^{2})^{-i}
\sum_{n=0}^{\infty}(mR)^{n}T^{(\textrm{int})}_{n,i}(s)\;,
\end{eqnarray}
where
\begin{eqnarray}\label{tees}
T^{(\textrm{int})}_{n,i}(s)&\!\!=\!\!&\sum_{l=0}^{\left[n/2\right]}\left[2\zeta_{H}\left(2s+i+2l-1,\frac{1}{2}\right)+\zeta_{H}\left(2s+i+2l,\frac{1}{2}\right)\right]\nonumber\\
&&\times\sum_{\sigma=\pm 1}\sum_{j=0}^{3i}d^{(\textrm{int})}_{n-2l,\sigma}(i,j)\Big[i\beta_{\sigma}\Lambda_{l}(s,i+j,i+1,\beta_{\sigma})-(i+j)\Lambda_{l}(s,i+j-1,i,\beta_{\sigma})\Big]\;.
\end{eqnarray} 
By exploiting the relations (\ref{aminus1ext}) and (\ref{a0ext}), the small-$m$ expansion for $A^{(\textrm{ext})}_{-1}(s)$ and for $A^{(\textrm{ext})}_{0}(s)$ can be immediately obtained from those for  $A^{(\textrm{int})}_{-1}(s)$ and $A^{(\textrm{int})}_{0}(s)$, respectively. For $A^{(\textrm{ext})}_{i}(s)$ 
we have, instead, 
\begin{eqnarray}\label{aifinalext}
A^{(\textrm{ext})}_{i}(s)&\!\!=\!\!&\frac{\sin(\pi s)}{2^{2i}\pi}R^{2s}(4+\gm^{2})^{-i}
\sum_{n=0}^{\infty}(mR)^{n}T^{(\textrm{ext})}_{n,i}(s)\;,
\end{eqnarray}
where $T^{(\textrm{ext})}_{n,i}(s)$ are obtained from (\ref{tees}) by replacing the coefficients $d^{(\textrm{int})}_{k,\sigma}(i,j)$ with $d^{(\textrm{ext})}_{k,\sigma}(i,j)$,
the latter being obtained from the relations for the exterior region in the Appendix \ref{app}.  
\end{proof}

\section{Casimir energy  in the confining case}
The analytic continuation (\ref{zetacontinued}) together with the explicit small-$m$ expansions (\ref{a-1final}), (\ref{a0final}), (\ref{b0final}), (\ref{aifinal}), and 
the corresponding ones for the exterior region, can now be used to compute the Casimir energy of a spinor field under the influence of a confining $\delta$-shell potential. 
In the framework of the spectral zeta function regularization the Casimir energy can be extracted by analyzing the zeta function of the system in a neighborhood of $s=-1/2$ \cite{kirsten2001spectral,bordag2009advances}. In the case of the Dirac spinor under the influence of the $\delta$-shell potential the expression for the Casimir energy is
\begin{equation}\label{casimirformula}
E_{\textrm{Cas}}=E_{\textrm{Cas}}^{\textrm{int}}+E_{\textrm{Cas}}^{\textrm{ext}}
=\lim_{\varepsilon\to 0}\frac{-\mu^{2\varepsilon}}{2}\left[\zeta^{(\textrm{int})}\left(\varepsilon-\frac{1}{2}\right)+\zeta^{(\textrm{ext})}\left(\varepsilon-\frac{1}{2}\right)\right]\;.
\end{equation}
As the reader might have noticed in the previous Section the spectral zeta function develops, in the process of analytic continuation, 
a simple pole at $s=-1/2$. This is a general feature of the zeta function regularization method \cite{kirsten2001spectral,bordag2009advances} and implies that 
the right-hand-side of (\ref{casimirformula}) has the small-$\varepsilon$ behavior
\begin{eqnarray}\label{casimirformula2}
&&\frac{-\mu^{2\varepsilon}}{2}\left[\zeta^{(\textrm{int})}\left(\varepsilon-\frac{1}{2}\right)+\zeta^{(\textrm{ext})}\left(\varepsilon-\frac{1}{2}\right)\right]\\
&=&-\frac{1}{2}\left(\frac{1}{\varepsilon}+\ln\mu^2\right)\left[\rs\zeta^{(\textrm{int})}\left(-\frac{1}{2}\right)+\rs\zeta^{(\textrm{ext})}\left(-\frac{1}{2}\right)\right]\nonumber\\
&-&\frac{1}{2}\left[\textrm{FP}\zeta^{(\textrm{int})}\left(-\frac{1}{2}\right)+\textrm{FP}\zeta^{(\textrm{ext})}\left(-\frac{1}{2}\right)\right]+O(\varepsilon)\;,\nonumber
\end{eqnarray}
where $\textrm{FP}\zeta$ denotes the finite part of the zeta function.
It is clear from the last formula that the Casimir energy of the system, given by the limit in (\ref{casimirformula}), is well-defined when the sum of the residues of the interior and exterior zeta functions at the point $s=-1/2$ vanishes. If this sum of the residues does not 
vanish, then the Casimir energy can be defined after a suitable renormalization of the theory. We refer to \cite{elizalde1998casimir} for a detailed description of the renormalization process. In order to select the configurations for which the Casimir energy is a well-defined quantity, we compute the total residue of the spectral zeta function at $s=-1/2$ and consider only those matching conditions that lead to a vanishing total residue. For a subset of these configurations, we then compute the finite part of the spectral zeta function at $s=-1/2$ which provides us with the Casimir energy of the spinor field. As a consistency check, throughout this section, we compare our findings for $\gm=-2$ with the ones found for MIT bag boundary conditions in \cite{elizalde1998casimir}.
We would like, at this point, to offer a few remarks regarding the coefficients of the heat kernel asymptotic expansion 
associated with our system. It is well known that the heat kernel coefficients provide the minimal set of counterterms necessary for the renormalization of a system in the ambit of quantum field theory \cite{kirsten2001spectral,vassilevich2003heat}. According 
to the general theory of the spectral zeta function (see e.g. \cite{kirsten2001spectral,vassilevich2003heat}), if $\mathcal{P}$ is a strongly elliptic second-order differential operator defined on a smooth compact Riemannian manifold with a smooth boundary and local boundary conditions then the coefficients $\mathcal{A}_{i}$ of the small-$t$ asymptotic expansion of $\textrm{Tr}e^{-t\mathcal{P}}$ can be extracted from the spectral zeta function $\zeta_{\mathcal{P}}(s)$ associated with the operator $\mathcal{P}$ according to the formulas
\begin{equation}\label{HKC1}
\mathcal{A}_{\frac{D}{2}-s}=\Gamma(s)\textrm{res}\zeta_{\mathcal{P}}(s)\;,
\end{equation}
with $s=(D-k)/2$, $k=\{0,\ldots,D-1\}$ and $s=-(2l+1/2)$, $l\in\mathbb{N}_{0}$, and 
\begin{equation}\label{HKC2}
\mathcal{A}_{\frac{D}{2}+p}=\frac{(-1)^{p}}{p!}\zeta_{\mathcal{P}}(-p)\;,\quad p\in\mathbb{N}_{0}\;.
\end{equation}
Since in the previous Sections we have derived the analytic continuation of the spectral zeta function associated with the spinor field 
propagating in $\mathbb{R}^{3}$ under the influence of the $\delta$-potential, we can use the last two formulas to discuss the heat kernel coefficients associated with our fermion system. It is important, however, to recall that a well-defined heat kernel expansion exists for suitable operators on smooth compact manifolds with or without a boundary. Our spinor field propagates, instead, in an unbounded space under the influence
of a $\delta$-potential. This means that by using the relations (\ref{HKC1}) and (\ref{HKC2}) with the \emph{total} spectral zeta function of our system, defined as the sum of the interior and exterior spectral zeta functions, we obtain \emph{relative} heat kernel coefficients, namely those defined as the difference between the heat kernel coefficients of our spinor system and those of the free spinor system in $\mathbb{R}^{3}$ (see e.g. \cite{fucci2016sphere} for a similar treatment in the case of scalar fields). With these comments in mind we can discuss the first few keat kernel coefficients.

In our particular case 
$D=3$, so the first heat kernel coefficient, namely $\mathcal{A}_{0}$, is proportional to the residue of the total spectral zeta function at $s=3/2$. Since the functions (\ref{zedint}) and (\ref{zedext}) are analytic for $\Re(s)>(3-N)/2$, the only possible contributions to the residue of the spectral zeta function, and hence to the heat kernel coefficients, come from the meromorphic functions $A^{(\textrm{int})}_{i}(s)+A^{(\textrm{ext})}_{i}(s)$. It is not difficult to see from the expressions for $A^{(\textrm{int})}_{i}(s)$ and $A^{(\textrm{ext})}_{i}(s)$ 
displayed in the previous Section, that only $A^{(\textrm{int})}_{-1}(s)$ and $A^{(\textrm{ext})}_{-1}(s)$ present a pole at $s=3/2$.
However, due to the relation (\ref{aminus1ext}) the total residue vanishes identically. This implies that $\mathcal{A}_{0}=0$, a
result that is to be expected since this relative coefficient is simply the difference between the volume of $\mathbb{R}^{3}$ with the potential and the volume of the free $\mathbb{R}^{3}$ which are obviously the same.

The next heat kernel coefficient, namely $\mathcal{A}_{1/2}$, is proportional to the residue of the total spectral zeta function at $s=1$.
It is easy to realize that a pole at this point only appears in the terms $A^{(\textrm{int})}_{0}(s)$ and $A^{(\textrm{ext})}_{0}(s)$. 
However, due to the relations (\ref{a0int}) and (\ref{a0ext}), we have that $A^{(\textrm{int})}_{0}(s)+A^{(\textrm{ext})}_{0}(s)=2b_{0}(s)$ where
$b_{0}(s)$ does not have a pole at $s=1$. This argument implies, then, that $\mathcal{A}_{1/2}=0$. The last result has a simple geometric interpretation: The coefficient $\mathcal{A}_{1/2}$ is expressed in terms of the curvature of the boundary. In our case the curvature of the sphere (which represents the boundary) has opposite sign when computed from the interior and exterior regions.
The computation of subsequent heat kernel coefficients $\mathcal{A}_{i}$ with $i\geq 1$ involves, in addition to the residues of $A^{(\textrm{int})}_{0}(s)$ and $A^{(\textrm{ext})}_{0}(s)$, also the residues of the functions $A^{(\textrm{int})}_{i}(s)$ and $A^{(\textrm{ext})}_{i}(s)$ with $i\geq 1$. As it is clear from the expressions in the previous Section, such computation is highly non-trivial and, therefore, cannot be added in detail here. However, we would like to remind the reader that the total residue of the spectral zeta function at the point $s=-1/2$ is proportional to the heat kernel coefficient $A_{2}$ which, when vanishing, indicates that the system has no 
ultraviolet divergences. In the next subsection we explicitly compute this residue, which we need in order to select the cases when the Casimir energy is a quantity needing no renormalization.

\subsection{Total residue at $\boldsymbol{s=-1/2}$}
We consider the divergent terms in the ground-state energy, which, as stated in \eqref{casimirformula2}, are proportional to the residue at $s=-1/2$ (equivalently $\varepsilon=0$) of the spectral zeta function. We can compute the interior and exterior contributions to the residue independently and define
\begin{equation}
\rs  A_{i'}\left(-\oh\right)\coloneqq\rs  A^{(\textrm{int})}_{i'}\left(-\oh\right)+\rs  A^{(\textrm{ext})}_{i'}\left(-\oh\right),\quad i'\in \{-1,0,1,2,3\}.
\end{equation}
Since all the residues are at $s=-1/2$, we will simply write $\rs  A_{i'}$.
From Eq.~\eqref{aminus1ext} it is easy to conclude that
\begin{equation}
\rs  A_{-1} =0.
\end{equation}
From Eqs.~\eqref{a0ext}, \eqref{a0final} and \eqref{b0final} we obtain
\begin{equation*}
\rs A_{0} = m^2 \left(R-\frac{\left(\gm^4+24 \gm^2+16\right) R}{8 \left(\gm^2+4\right) | \gm| }\right)\;.
\end{equation*}
For the remaining indices, $i = 1,2,3$, we use the relation \eqref{aifinal}. More explicitly, for $A_1(s)$ we have
\begin{eqnarray*}
\rs A_{1}&=&-\frac{2 \gm m}{3 \pi  \gm^2+12 \pi }-\frac{m^2 R \left(-4 | \gm| +\gm^2+4\right)}{8 | \gm| }\\[1ex]
&+&\frac{\left(\gm^2+4\right) m^3 R^2 \left[4 | \gm| +\left(\gm^2-4\right) \sin ^{-1}\left(\frac{\gm^2-4}{\gm^2+4}\right)\right]}{8 \pi  \gm | \gm| },
\end{eqnarray*}
where there are some cancellations between the interior and exterior contributions, which also occur for $i = 2,3$:
\begin{eqnarray*}
	\rs A_{2}&=&-\frac{1024 \left(\gm^2+4\right) | \gm| ^3+\gm^8-144 \gm^6-2976 \gm^4-2304 \gm^2+256}{64 \left(\gm^2-4\right)^4 R}\\[0.5ex]
	&+&\frac{4 \gm m}{\pi  \gm^2+4 \pi }-\frac{\left(\gm^2+4\right) m^2 R}{8 | \gm| },\\[1ex]
\hspace{-.8cm}	\rs A_{3}&\!\!\!\!=\!\!\!\!&\frac{1024 \left(\gm^2+4\right) | \gm| ^3+\gm^8-144 \gm^6-2976 \gm^4-2304 \gm^2+256}{64 \left(\gm^2-4\right)^4 R}\\[0.5ex]
	&\!\!\!\!+\!\!\!\!& \frac{m}{3 \pi  \left(\gm^2-4\right)^5 \left(\gm^2+4\right)}\biggl[48 \gm \left(\gm^2+4\right)^2 \left(\gm^2+12\right) \left(3 \gm^2+4\right) | \gm|  \sin ^{-1}\left(\frac{\gm^2-4}{\gm^2+4}\right)  \\[0.5ex]
	 &\!\!\!\!-\!\!\!\!& 2 \gm (\gm^{2}-4)\left(13 \gm^8+336 \gm^6+3040 \gm^4+5376 \gm^2+3328\right)\biggr]+\frac{\left(\gm^2+4\right) m^2 R}{8 | \gm| }
	\\[1ex]
	&\!\!\!\!-\!\!\!\!& \frac{m^3 R^2 \left[3 \left(\gm^2-4\right) \left(\gm^2+4\right)^2 | \gm|  \sin ^{-1}\left(\frac{\gm^2-4}{\gm^2+4}\right)+4 \left(3 \gm^4+8 \gm^2+48\right) \gm^2\right]}{24 \pi  \gm^3 \left(\gm^2+4\right)}.
\end{eqnarray*}
In all cases, our residues for $\gm = -2$ are in agreement with those found in \cite{elizalde1998casimir}. For the purpose of selecting those configurations for which the Casimir energy is well-defined,
we are interested in the sum of the residues, i.e.,
\begin{equation}
\rs(\gm,m)\coloneqq \sum_{i'=-1}^3 \rs A_{i'}\;,
\end{equation}
which can be found to be
\begin{eqnarray}\label{totalres}
\rs(\gm,m)&=&\frac{2 m}{3 \pi  \left(\gm^2-4\right)^5 \left(\gm^2+4\right)}\biggl[ 24 \gm \left(\gm^2+4\right)^2 \left(\gm^2+12\right) \left(3 \gm^2+4\right) | \gm|  \sin ^{-1}\left(\frac{\gm^2-4}{\gm^2+4}\right) \nonumber \\
&-& 8 \gm (\gm^{2}-4) \left(\gm^8+52 \gm^6+320 \gm^4+832 \gm^2+256\right)\biggr]\nonumber\\[1ex]
&-&\frac{m^2 R \left[-6 \left(\gm^2+4\right) | \gm| +\gm^4+16 \gm^2+16\right]}{4 \left(\gm^2+4\right) | \gm| }+\frac{8 \gm m^3 R^2}{3 \pi  \gm^2+12 \pi }.
\end{eqnarray}
By setting $\gm=-2$ in the previous expression, we obtain the total residue in the MIT case
\begin{equation}
\rs(\gm=-2,m)=-\frac{2 m^3 R^2}{3 \pi }-\frac{2 m}{15 \pi },
\end{equation}
in accordance with the result found in \cite{elizalde1998casimir}.

It is important to make a remark at this point. The explicit expression for the total 
residue in \eqref{totalres} shows that for a massless field $\rs(\gm,m=0)=0$ regardless 
of the value of $\gm$. This implies that the Casimir energy in the massless case is always well-defined. For a massive field, instead, $\rs(\gm,m)$ does not, in general, vanish. 

For instance, in the case of MIT bag boundary conditions there are no positive values of $m$ such that the residue is zero. As we have already mentioned before, the Casimir energy in this case can be defined only after a suitable renormalization process as described in e.g. \cite{elizalde1998casimir}. Even after a renormalization procedure, however, there are still parameters which need to be determined experimentally. As explained in \cite{elizalde1998casimir,bordag2009advances} the experimental determination of the necessary parameters cannot be achieved within the model since it is not yet suitable for describing a realistic physical situation. In other words, more study is needed in order to obtain physically meaningful results for the Casimir energy of massive spinor fields in the MIT case.

\begin{figure}[h!]
	\centering
	\includegraphics[width=.65\textwidth]{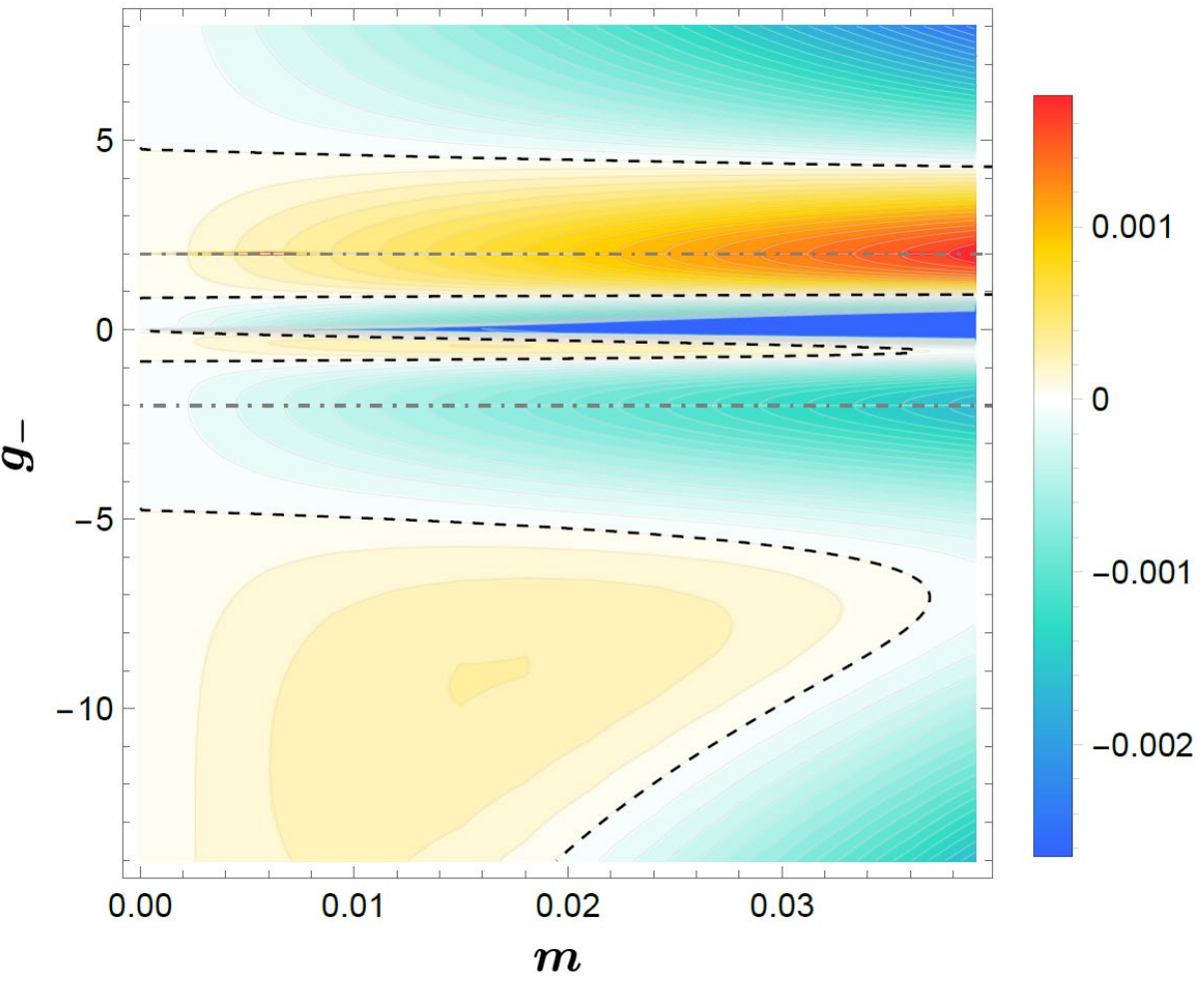}
	\caption{Contour plot of the function $\rs(\gm,m)$ for $R=1$. The black dashed curve is defined by $\rs(\gm,m)=0$. Two gray dot-dashed lines are plotted at $\gm=\pm 2$, where $\gm=- 2$ leads to MIT bag boundary conditions.}   
	\label{fig:1}
\end{figure}

By considering more general matching conditions, as we have done in this paper, we 
are able to describe a much wider set of models of spinor fields confined within a spherical shell. In particular, the expression for the total residue in \eqref{totalres} represents one of the main results of our work and shows that there exist confinement models of massive spinor fields for which the Casimir energy can be well-defined \emph{without} the need for renormalization. In Fig.~\ref{fig:1} we obtained a plot of the total residue for $R=1$ in terms of the mass of the spinor field and the parameter $\gm$ characterizing the matching conditions. The black dotted lines represent the curves in the $(m,\gm)$-plane where the
residue is identically zero. It is clear from this plot that for a given mass of the spinor field we can find at least one value of $\gm$ for which the residue vanishes. In other words, for a given massive spinor field propagating within a $\delta$-shell potential of fixed radius one can tune the matching conditions to obtain a well-defined Casimir energy.
The existence of at least one value of $\gm$ for which the residue vanishes for fixed $m$ and $R$ can be proved by noticing that for $\gm\to 2$, $\rs(2,m)=-\rs(-2,m)>0$  and that 
\begin{equation}
\lim_{\gm\to\pm\infty}\rs(\gm,m)=-\infty\;,
\end{equation}  
for any given $m$ and $R$.
Since $\rs(\gm,m)$ is a continuous functions for $\gm\neq 0$, there exists at least one value in the interval $[2,+\infty)$ for which $\rs(\gm,m)=0$.   

From the same plot one can realize that if one fixes $R$ and $\gm$, that is a specific $\delta$-shell potential is considered, one cannot always find a massive spinor field for which the total residue vanishes and, hence, the Casimir energy can be defined without renormalization. This can be easily understood by analyzing the expression \eqref{totalres}.
By fixing the values of $R$ and $\gm$ the total residue becomes a polynomial in $m$ of degree 3. While three roots of this polynomial always exist they are not guaranteed to be positive.   

The above remarks can be summarized as follows: while for a given massive spinor field we can 
always find a specific $\delta$-shell potential for which the Casimir energy does not need renormalization, the reverse is not always true. 

\subsection{Casimir energy}

For those configurations in which the total residue vanishes, the Casimir energy of the system is simply given by the finite part of the spectral zeta function at $s=-1/2$.  
In the rest of the paper we focus on massless spinor fields mainly for three reasons: first, the energy in this case is well-defined for all values of the parameter $\gm$. This allows us to consider a wide variety of matching conditions without having to deal with issues involving renormalization. Second, the numerical evaluations required for the computation of the Casimir energy are much less demanding. Lastly, an explicit expression for the finite part of the zeta function which is valid for all values of the mass cannot, in general, be obtained. One can, however, derive expressions that are valid for small values of the mass (see e.g. \cite{elizalde1998casimir}). 
Restricting our analysis to the massless field, however, is not deleterious. In fact, we will show that the massless spinor field in $\delta$-shell potentials displays very interesting features such as both attractive and repulsive pressures on the shell. 

According to \eqref{zetacontinued}, we have two contributions to the finite part of the zeta function at $s=-1/2$. The first, coming from the zero order terms of the series of $A_{i'}(\varepsilon-1/2)$ around $\varepsilon=0$ and the other coming from the integral $Z^{(\textrm{int}/\textrm{int})}(-1/2)$.
In order to evaluate the contributions coming from the terms $A_{i'}(\varepsilon-1/2)$, we first define 
\begin{equation}
A_{i'}(s)\coloneqq A^{(\textrm{int})}_{i'}(s)+A^{(\textrm{ext})}_{i'}(s),\quad i'\in \{-1,0,1,2,3\}.
\end{equation}
For $i'=-1$, relation \eqref{aminus1ext} ensures that $A_{-1}(-1/2)=0$. From  Eqs.~\eqref{a0ext}, \eqref{a0final} and \eqref{b0final}, in the limit $\varepsilon\to 0$, we find
\begin{equation}
\textrm{FP}A_0=\frac{-4 | \gm| +\gm^2+4}{12 \gm^2 R+48 R}.
\end{equation}
By utilizing the expression in \eqref{aifinal}, similar results can be found for $\textrm{FP}A_i'$ for $i'=\{1,2,3\}$. Specifically, we find
\begin{eqnarray*}
 \textrm{FP}A_1&\!\!\!\!=\!\!\!\!& -\textrm{FP}A_0,\\[1ex]
  \textrm{FP}A_2&\!\!\!\!=\!\!\!\!&-\frac{1}{64 \left(\gm^2-4\right)^4 \left(\gm^2+4\right) R}\Big[1024 \left(\gm^2+4\right)^2 | \gm| ^3 \log \left(\left(\gm^2+4\right) R^2\right)\\
  &\!\!\!\!+\!\!\!\!&2048 \left(\gm^2+4\right)^2 | \gm| ^3 \coth ^{-1}\left(\frac{1}{2} \left(\gm^2+2\right)\right)+\frac{2 \gamma  \left(\gm^2+4\right)}{(| \gm| -2)^{-4}} \left(8 \left(\gm^2+4\right) | \gm| +\gm^4-104 \gm^2+16\right)\\
  &\!\!\!\!-\!\!\!\!&2 (| \gm| -2)^2 \left[\left(\gm^2-4\right)^2 \left(13 \gm^4-24 \gm^2+208\right)-4 \left(\gm^2+4\right) \left(3 \gm^4-280 \gm^2+48\right) | \gm| \right]\\
  &\!\!\!\!+\!\!\!\!&2 \gm^{10} \log (4 R)-8 \left(\gm^2+4\right) \left(35 \gm^4+748 \gm^2+560\right) \gm^2 \log (4 R)+2048 \log (R)+4096 \log (2)\Big],\\[1ex]
  \textrm{FP}A_3&\!\!\!\!=\!\!\!\!& \frac{1}{384 \left(\gm^2-4\right)^4 \left(\gm^2+4\right) R}\left(3 \pi ^2 \left(\gm^2+4\right) \left(| \gm|  \left(| \gm|  \left(8 | \gm| +\gm^2-104\right)+32\right)+16\right) (| \gm| -2)^4 \right. \\
  &\!\!\!\!+\!\!\!\!& \left. 12 \gamma  \left(\gm^2+4\right) \left(| \gm|  \left(| \gm|  \left(8 | \gm| +\gm^2-104\right)+32\right)+16\right) (| \gm| -2)^4\right. \\
  &\!\!\!\!-\!\!\!\!& \left. 16 \left(\gm^2-4\right)^2 \left(| \gm|  \left(| \gm|  \left(-32 \left(\gm^2+4\right) | \gm| +5 \gm^4+172 \gm^2+688\right)-512\right)+320\right)\right. \\
  &\!\!\!\!+\!\!\!\!& \left.12 \left(\gm^2+4\right) \left(1024 \left(\gm^2+4\right) | \gm| ^3 \log \left(\left(\gm^2+4\right) R\right)+\left(\gm^8-144 \gm^6-2976 \gm^4-2304 \gm^2+256\right) \log (R)\right)\right. \\
  &\!\!\!\!+\!\!\!\!& \left.24 \left(\gm^2+4\right) \left(\gm^8-144 \gm^6-2976 \gm^4-2304 \gm^2+256\right) \log (2)\right)-12288 \left(\gm^2+4\right)^2 | \gm| ^3 \log (| \gm| ),
\end{eqnarray*}
where $\gamma$ is Euler's constant. All of these expressions are simpler for MIT bag boundary conditions. In fact, in this case they read
\begin{eqnarray*}
	\textrm{FP}A_0&\!\!\!\!=\!\!\!\!&\textrm{FP}A_1= 0,\\[1ex]
	\textrm{FP}A_2&\!\!\!\!=\!\!\!\!& \frac{\log (R)+\gamma -1+\log (4)}{32 R},\\[1ex]
	\textrm{FP}A_3&\!\!\!\!=\!\!\!\!& -\frac{4 \log (R)+\pi ^2+4 \gamma +\log (256)}{128 R},
\end{eqnarray*}
and are in agreement with the numerical values for $R=1$ provided in Table 2 of \cite{elizalde1998casimir}.
The only task left consists in the computation of $Z^{(\textrm{int})}(s=-1/2)$ and $Z^{(\textrm{ext})}(s=-1/2)$ in Eqs.~\eqref{zedint} and \eqref{zedext}, respectively. The integration required for this calculation can only be performed numerically. To this end, it is convenient to integrate by parts, to obtain
\begin{eqnarray}\label{intbyparts}
	Z^{(\textrm{int/ext})}(-1/2)&=&-\frac{1}{\pi}
	\sum^{\infty}_{j=1/2,3/2,\cdots}(2j+1)\sum_{\sigma=\pm 1}\int_{\frac{m R}{j}}^{\infty}\mathrm{d}z\left[\left(\frac{zj}{R}\right)^{2}-m^{2}\right]^{1/2}\nonumber \frac{\partial}{\partial z}{F}^{(\textrm{int/ext})}_{j,\sigma}(z)\\
	&=& \frac{1}{\pi}
	\sum^{\infty}_{j=1/2,3/2,\cdots}(2j+1)\sum_{\sigma=\pm 1}\int_{\frac{m R}{j}}^{\infty}\mathrm{d}z\frac{z j/R}{\sqrt{z^2-\frac{m^2 R^2}{j^2}}}\nonumber {F}^{(\textrm{int/ext})}_{j,\sigma}(z),
\end{eqnarray}
where the function ${F}^{(\textrm{int/ext})}_{j,\sigma}(z)$ is defined according to Eqs.~\eqref{zedint} and \eqref{zedext}. Note that the boundary terms from the integration by parts vanish. In fact, $[(zj/R)^{2}-m^{2}]^{1/2}{F}^{(\textrm{int/ext})}_{j,\sigma}(z)\to 0$ both as $z\to mR/j$ and as $z\to\infty$. The latter is due to the fact that ${F}^{(\textrm{int/ext})}_{j,\sigma}(z)\to 0$ 
as $z\to\infty$. 

By using the formula \eqref{casimirformula2} and by recalling that $\textrm{FP}A_{-1}=0$ and that $\textrm{FP}A_0=-\textrm{FP}A_1$, we obtain the following expression for the renormalized Casimir energy
\begin{equation}\label{E0ren}
E_{\textrm{Cas}}^\text{ren}=-\oh\left(\textrm{FP}A_2+\textrm{FP}A_3+Z(-1/2)\right),
\end{equation}
where $Z(-1/2)$ denotes the sum of the interior $Z^{(\textrm{int})}(-1/2)$ and exterior $Z^{(\textrm{ext})}(-1/2)$ contributions. We emphasize that we do not need a classical model in order to accommodate the renormalization \cite{elizalde1998casimir,bordag2009advances} since in the massless case considered here the total residue of the zeta function is identically zero. 

\begin{figure}[h!]
	\centering
	\includegraphics[width=.7\textwidth]{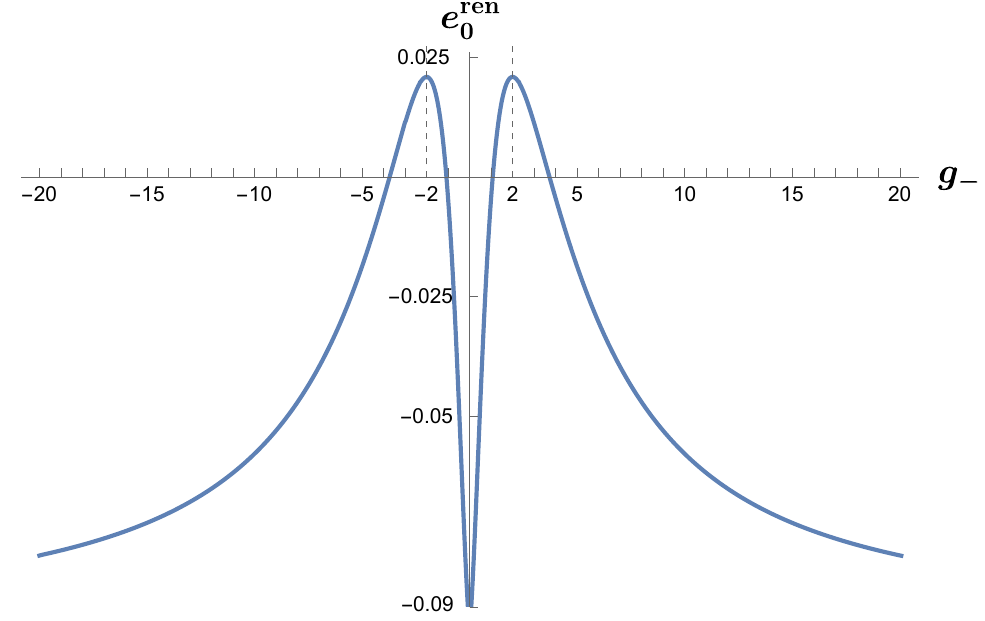}
	\caption{Renormalized energy for the fermionic massless field for $R=1$, $E_\text{Cas}^\text{ren}={e_0^\text{ren}}/{R}$, as a function of $\gm$.}   
	\label{fig:2}
\end{figure}

We would like to point out that by following a dimensional analysis, the Casimir energy in \eqref{E0ren} can explicitly written as
\begin{equation}\label{E0ren1}
E_{\textrm{Cas}}^\text{ren}=\frac{e_0^\text{ren}}{R}.
\end{equation}
This can be justified as follows: Since we have assumed that $\hbar=1$ and $c=1$, we can write $[E_{\textrm{Cas}}^\text{ren}]=[m]=\text{L}^{-1}$. From Eq.~\eqref{eq:Potential}, and noting that
$\delta(r-R)=R^{-1}\delta(r/R-1)$, we can conclude that $[g_v]=[g_s]=\emptyset$. These remark imply that the only parameter with dimensions in our system is the radius $R$, and, therefore, \eqref{E0ren} represents the only possible $R$-dependence satisfying $[E_\text{Cas}^\text{ren}]=\text{L}^{-1}$.
The plot displayed in Fig.~\ref{fig:2} shows the dependence of $e_0^\text{ren}$ on the parameter $\gm$.
One can immediately notice that the energy is invariant with respect to the change $\gm\to -\gm$. This follows from the symmetry of the characteristic functions \eqref{eq:mode_gen} and \eqref{2022} under this transformation in the massless case. This symmetry, however, does not hold in the massive case as one can discern from the characteristic functions.
Note that the case $\gm = 0$ is excluded since the confining condition $g_v^2-g_s^2+4=0$ can not be satisfied if $g_s=g_v$. However, we can consider values such that $\gm\to 0$, which leads to $|g_v|, |g_s| \to \infty$.
We would like to mention that the result for MIT bag boundary conditions computed in \cite{elizalde1998casimir}, that is,
\begin{equation}
E_\text{Cas}^\text{ren}\simeq \frac{0.0204}{R},
\end{equation}
is recovered from our results. More precisely, by using Newton's method it can be shown that the maximum value of the energy is attained at $g_{-}=\pm 2$. This implies, in particular, that MIT boundary conditions generate the largest value for the Casimir energy. This situation is similar to that of a scalar field propagating under the presence of a singular interaction analogous to the one considered in Eq.~\eqref{eq:Potential}, the so-called $\delta$-$\delta'$ interaction \cite{romaniega2022casimir}. In the paper \cite{romaniega2022casimir} it was found that the maximum value of the energy occurs when this interaction approaches Dirichlet and Robin boundary conditions. In a sense, then, the MIT boundary conditions play, in our work, the same role that Dirichlet boundary conditions do in the $\delta$-$\delta'$ interaction for scalar fields. A direct comparison is, however, not possible since Dirichlet boundary conditions do not lead to a self-adjoint Dirac Hamiltonian \cite{alonso1997boundary}.

The most important observation of this section, however, relates to the sign of the 
Casimir energy. From the plot of the energy in Fig.~\ref{fig:2} one realizes that there 
are values of the parameter $\gm$ for which the Casimir energy becomes negative. 
This is a completely novel feature that does not appear in the MIT bag model. In the MIT 
case, in fact, the energy is always positive. This property leads to rather interesting situations. 
For instance, the pressure acting on the spherical shell can be derived from the renormalized energy using the principle of virtual work \cite{barton2004casimir,li2019casimir}. For our system we have, in particular, 
\begin{equation}\label{eq:MeanPress}
	p^\text{ren}_\text{Cas}= -\dfrac{1}{4\pi R^{2}}\frac{\partial  E_\text{Cas}^\text{ren}}{\partial  R} = \frac{e_0^\text{ren}}{4\pi R^4}.
\end{equation}
This relation shows that in the MIT case only an outward pressure appears \cite{milton2001casimir} since the Casimir energy is always positive. In our case, however, since the Casimir energy can be both positive and negative for different values of $\gm$, the system can exert both inward and outward pressure on the spherical shell. Moreover, 
the amount of pressure acting on the shell changes in magnitude depending on the value 
of $\gm$ (i.e. for different types of matching, or boundary, conditions).
Another interesting observation is the following: since the pressure changes sign, there are values of $\gm$ for which the pressure vanishes (as it can be seen in Fig.~\ref{fig:2}). 
This means that by tuning the matching conditions, one can obtain a fermionic system that generates \emph{no pressure} on the spherical shell. From the graph in Fig.~\ref{fig:2}, however, it is clear that such configuration is unstable since a small variation of $\gm$ to either side of the equilibrium value would give rise to either an outward or inward pressure.
Nevertheless, this is an intriguing phenomenon which, to our knowledge, has not been previously observed in other spherically symmetric spinor systems.

We conclude this section by noting that the values for the Casimir energy found here are of the same order of magnitude as the ones found for the electromagnetic field in the presence of a perfectly conducting shell \cite{boyer1968quantum}
\begin{equation}\label{eq:ErenEM}
E_{\text{Cas}, \text{EM}}^\text{ren}\simeq \frac{0.0461766}{R}.
\end{equation}
However, for the scalar field  the values for a Dirichlet and a Neumann sphere are $E_{\text{Cas}, \text{D}}^\text{ren}\simeq 0.0028168/R$ and $E_{\text{Cas}, \text{N}}^\text{ren}\simeq -0.2238216/R$, respectively \cite{bordag2009advances}, being also $E_{\text{Cas}, \delta\textit{-}\delta'}^\text{ren}\simeq 0.001/R\geq 0$ for the $\delta$-$\delta'$ aforementioned interaction \cite{romaniega2022casimir}.

\section{Conclusions}

In this paper we have analyzed the Casimir energy of a spin one-half field around classical configurations which generalize the MIT bag model.
In particular, we have studied the spectral zeta function associated with a spinor field propagating in $\mathbb{R}^{3}$ under the influence of a singular potential constructed as a combination of a scalar and vector $\delta$-potentials supported on a spherical shell. The definition of this potential is based on the theory of self-adjoint extensions, and different values of the scalar and vector couplings lead to self-adjoint extensions that describe either a confining or non-confining system. We have then performed the analytic continuation of the corresponding spectral zeta function which we used to analyze the Casimir energy of a massless spinor field in the confining case.
We have found that the mathematical machinery needed for the analysis of the spectral zeta function is much more involved than both the one used to treat a scalar field under the influence of a $\delta$-shell potential and the one utilized to study the MIT bag model. 
This is due, in part, by the presence of the vector potential.

Nevertheless, the additional technical complications needed in our analysis are a price well-worth paying since we have found that the Casimir energy of the spinor field obeying matching conditions determined by a $\delta$-shell potential displays interesting characteristics that have not been previously observed in other spinor systems. 

As the first noteworthy result of this work, we have shown that for a given mass of the spinor field and radius of the spherical shell we have at least one value of $\gm$ such that the total residue vanishes identically. In this case, then, the vacuum energy is well-defined without the need for any renormalization procedure. This is important since the renormalization process for massive spinor fields in spherically symmetric systems might not lead to meaningful results. For instance, a well-known renormalization procedure in quantum field theory relies on the small-$t$ expansion of the trace of the heat kernel associated with the spectral function of the system. This expansion provides the large-$m$ behavior of the theory. The divergent terms, as $m\to\infty$ (the classical limit), of the expansion provide the counterterms needed to renormalize the physical parameters of the system. In the massive spinor case, the large-mass behavior of the theory is not simply given by the heat kernel asymptotic expansion due to the explicit dependence of the characteristic functions on the mass $m$. For this reason the implementation of the large-mass normalization condition cannot be achieved analytically \cite{kirsten2001spectral}. Other renormalization processes also encounter difficulties since they cannot determine all the parameters of the theory \cite{elizalde1998casimir}. Specifically, in the latter when both the interior and exterior regions are considered, two parameters, which cannot be theoretically determined, still remain in the model. These parameters must be adjusted numerically  by comparing them directly to the actual physical system described. However, the model, as it stands, cannot be considered a realistic description of a physical system.
This situation is similar to the one in which boundary conditions modeling a perfect conductor are imposed on the electromagnetic field. The perfect conductor boundary conditions are based on the notion that electrons flow on the surface in such a way that any incoming field is annihilated. However, this description still represents an idealization, since the effect of imperfect reflections is known to be large in experiments \cite{dalvit2011casimir}. Therefore, those parameters that cannot be fixed by the renormalization process need to be found experimentally. 
Unfortunately, very few systems exist in which the Dirac Hamiltonian is exactly solvable and, consequently, opportunities for an experimental determination of the unknown parameters are rare. 
In these circumstances, point-supported potentials, like the $\delta$-shell and the MIT bag, as one of its particular cases, become both theoretically and physically relevant: in fact, they are simple enough to be exactly solvable \cite{albeverio2005solvable} but involved enough to be able to describe idealized physical systems. Indeed, as already mentioned, within our model we are able to find configurations for which the above-mentioned problem regarding a meaningful renormalization does not arise. For instance, the $\delta$-shell potential defined in (\ref{eq:Potential}) can be used to model a class of \emph{impenetrable} short-range potentials by varying the relative size of $g_s$ and $g_v$ within the constraint (\ref{constraint}). Moreover, Dirac equation endowed with point supported potentials are particularly important in the description of two dimensional systems such as graphene in the field of an Aharonov-Bohm solenoid orthogonal to its plane \cite{jakiw2009}, magnetic Kronig-Penney model for Dirac electrons in single-layer graphene \cite{masir2009}, and other models of propagation of fermions in the presence of impurities \cite{guilarte2019one}. The results of this paper are, therefore, relevant
for the analysis of the Casimir energy of fermions in the aforementioned models.
 
The second important result of our paper consists in the computation of the energy and pressure acting on the surface of the spherical shell in the case of a massless spinor field. We have found that, for a given radius of the spherical shell, the Casimir energy varies both in magnitude and sign as the parameter $\gm$ changes. This is an important observation since by changing the matching conditions one can transition from an outward to an inward Casimir pressure on the shell. This appears to be a novel feature   
since in the MIT bag model only an outward pressure was found, like in Boyer's result for a perfectly conducting sphere. The case presented here is relevant since it preserves confinement, which is a rough representation of asymptotic freedom and the main reason why the MIT model is widely used. However, unlike the MIT bag model, confinement is achieved not only with a large set of values for the pressure but also with both inward and outward pressure. These results could be useful within the context of bag models, where there are some unresolved problems regarding the sign and magnitude of the pressure from the zero-point energy of the quark field \cite{milton2001casimir}. 

We would like to conclude this Section by mentioning some natural continuations of this research. This paper is focused on matching conditions leading to confinement. It would be particularly interesting to complement this analysis by considering the Casimir energy of massless spinor fields propagating in a space where the $\delta$-shell potential leads to non-confining configurations. Any result in this area 
would be interesting as these types of $\delta$-potentials would model semi-transparent bags. 
Our analysis of the spectral zeta function and the Casimir energy has been conducted for massive scalar field but it would be very interesting to extend it to gauge fields. In particular, one could generalize the analysis performed in, for instance, \cite{wipf95} regarding gauge fields in the MIT bag to gauge fields propagating in a flat space under the influence of a $\delta$-potential of the type considered in this paper. This investigation would not only provide the relevant spectral zeta function and Casimr energy but would also give details related to the behavior of the $\delta$-potential under gauge transformations. In particular one could repeat the analysis performed in this paper for the Dirac Hamiltonian $H_0 = -i \alpha_i \nabla_i + m\beta$ with $\nabla_{j}=\partial_j+iA_{j}$ and study the interaction of the gauge fields $A_{j}$ with the $\delta$-potential. We hope to report on these interesting topics in future works.    

\section{Acknowledgments}

C\'esar Romaniega is grateful to the Spanish Government for funding under the FPU-fellowships program FPU17/01475, the FPU mobility program  EST21/00286 and MCIN grant  PID2020-113406GB-I00.

\appendix

\section{Asymptotic expansion of $\mathcal{G}^{(\textrm{int})}_{j,\sigma}$ and $\mathcal{G}^{(\textrm{ext})}_{j,\sigma}$}\label{app}

By using the uniform asymptotic expansion for the modified Bessel functions and their first derivative which are provided in Chapter 10, Section 41 of \cite{olver2010nist}, we can obtain the following expressions
\begin{eqnarray}\label{app1}
I^{2}_{j}(zj)&\!\!=\!\!&\frac{e^{2j\eta}}{(2\pi j)(1+z^2)^{1/2}}\sum_{k=0}^{\infty}\frac{u_{k}(t)}{j^{k}}\;,\nonumber\\
(I')^{2}_{j}(zj)&\!\!=\!\!&\frac{e^{2j\eta}}{(2\pi j)z^{2}}(1+z^2)^{1/2}\sum_{k=0}^{\infty}\frac{v_{k}(t)}{j^{k}}\;,\nonumber\\
I_{j}(zj)I'_{j}(zj)&\!\!=\!\!&\frac{e^{2j\eta}}{(2\pi j)z}\sum_{k=0}^{\infty}\frac{m_{k}(t)}{j^{k}}\;,
\end{eqnarray}
where
\begin{eqnarray}
\eta=(1+z^2)^{1/2}+\ln\left(\frac{z}{1+(1+z^2)^{1/2}}\right)\;,\quad t=(1+z^2)^{-1/2}\;,
\end{eqnarray}  
and
\begin{eqnarray}
u_{k}(t)&\!\!=\!\!&\sum_{l=0}^{k}U_{l}(t)U_{k-l}(t)\;,\quad v_{k}(t)=\sum_{l=0}^{k}V_{l}(t)V_{k-l}(t)\;,\nonumber\\
m_{k}(t)&\!\!=\!\!&\sum_{l=0}^{k}U_{l}(t)V_{k-l}(t)\;, 
\end{eqnarray}
with the functions $U_{k}(t)$ and $V_{k}(t)$ given by the recurrence relations
\begin{eqnarray}
U_{k+1}(t)&\!\!=\!\!&\frac{1}{2}t^{2}(1-t^{2})U'_{k}(t)+\frac{1}{8}\int_{0}^{t}(1-5\tau^{2})U_{k}(\tau){\dtau}\;.\nonumber\\
V_{k+1}(t)&\!\!=\!\!&U_{k+1}(t)-\frac{1}{2}t(1-t^{2})U_{k}(t)-t^{2}(1-t^{2})U'_{k}(t)\;.
\end{eqnarray}
For the terms involving the modified bessel functions of the second kind we have instead
\begin{eqnarray}\label{app2}
K^{2}_{j}(zj)&\!\!=\!\!&\left(\frac{\pi}{2j}\right)\frac{e^{-2j\eta}}{(1+z^2)^{1/2}}\sum_{k=0}^{\infty}\frac{(-1)^{k}u_{k}(p)}{j^{k}}\;,\nonumber\\
(K')^{2}_{j}(zj)&\!\!=\!\!&\left(\frac{\pi}{2j}\right)\frac{e^{-2j\eta}}{z^{2}}(1+z^2)^{1/2}\sum_{k=0}^{\infty}\frac{(-1)^{k}v_{k}(p)}{j^{k}}\;,\nonumber\\
K_{j}(zj)K'_{j}(zj)&\!\!=\!\!&-\left(\frac{\pi}{2j}\right)\frac{e^{-2j\eta}}{z}\sum_{k=0}^{\infty}\frac{(-1)^{k}m_{k}(p)}{j^{k}}\;.
\end{eqnarray}
The expansions (\ref{app1}) employed in the expression for the characteristic function (\ref{modegenint})
allows us to obtain
\begin{eqnarray}
\mathcal{G}^{(\textrm{int})}_{j,\sigma}(z)=\frac{e^{2j\eta}}{\pi j z^{2j+2}}(1+z^2)^{1/2}(1-t)\sum_{k=0}^{\infty}\frac{\Omega^{(\textrm{int})}_{k,\sigma}(t)}{j^{k}}\;,
\end{eqnarray}
where we find $\Omega^{(\textrm{int})}_{0,\sigma}(t)$ to be
\begin{eqnarray}\label{omegazeroint}
\Omega^{(\textrm{int})}_{0,\sigma}(t)=\frac{1}{4}\left(4+\gm^{2}\right)
+\frac{\sigma}{4}\left(4-\gm^{2}\right)t\;,
\end{eqnarray}
and $\Omega_{k,\sigma}(t)$, with $k\geq 1$,
\begin{eqnarray}
\Omega^{(\textrm{int})}_{k,\sigma}(t)&\!\!=\!\!&\left[\left((1-t^{2})\frac{\gm^{2}}{4}+t^{2}\right)
\sigma_{-}
+\left(1-t^{2}+\frac{\gm^{2}}{4}t^{2}\right)
\sigma_{+}\right]\frac{u_{k}(t)}{1-t}\\
&+&\left[
\sigma_{-}+
\frac{\gm^{2}}{4}\sigma_{+}
\right]\frac{v_{k}(t)-2tm_{k}(t)}{1-t}-\frac{mR\gm t}{1-t}[m_{k-1}(t)-tu_{k-1}(t)]\;.\nonumber
\end{eqnarray}
In a similar fashion, by using the expansions (\ref{app2}) in the characteristic function (\ref{modegenext})
we get
\begin{eqnarray}
\mathcal{G}^{(\textrm{ext})}_{j,\sigma}(z)=\frac{\pi e^{-2j\eta}}{j z^{-2j}}(1+z^2)^{1/2}(1+t)\sum_{k=0}^{\infty}\frac{\Omega^{(\textrm{ext})}_{k,\sigma}(t)}{j^{k}}\;,
\end{eqnarray}
where one finds
\begin{eqnarray}\label{omegazeroext}
\Omega^{(\textrm{ext})}_{0,\sigma}(t)=\Omega^{(\textrm{int})}_{0,-\sigma}(t)\;,
\end{eqnarray}
and, for $k\geq 1$,
\begin{eqnarray}
\Omega^{(\textrm{ext})}_{k,\sigma}(t)&=&\frac{(-1)^{k}}{1+t}\Bigg\{\left[\left((1-t^{2})\frac{\gm^{2}}{4}+t^{2}\right)
\sigma_{-}
+\left(1-t^{2}+\frac{\gm^{2}}{4}t^{2}\right)
\sigma_{+}\right]u_{k}(t)\nonumber\\
&+&\left[
\sigma_{-}+
\frac{\gm^{2}}{4}\sigma_{+}
\right][v_{k}(t)+2tm_{k}(t)-(mR\gm t)[m_{k-1}(t)+tu_{k-1}(t)]\Bigg\}\;.\nonumber\\
\end{eqnarray}
For the purpose of analytic continuation we need the functions $\mathcal{D}^{(\textrm{int})}_{i,\sigma}(t)$ and $\mathcal{D}^{(\textrm{ext})}_{i,\sigma}(t)$ expressed through the expansions (\ref{cumulint}) and (\ref{cumulext}), respectively. In particular, in order to evaluate the Casimir energy we only need the first three. With the help of an algebraic computer program and by noticing that 
\begin{eqnarray}
\left(\sigma_{\pm}\right)^{n}&\!\!=\!\!&\sigma_{\pm}\;,\quad n\in {\mathbb{N}_{0}}\;,\nonumber\\
\sigma_{+}\sigma_{-}=0\;,\quad \sigma_{+}-\sigma_{-}&\!\!=\!\!&\sigma\;,\quad \sigma_{+}+\sigma_{-}=1\;,
\end{eqnarray}
we obtain
\begin{small}
{\begin{eqnarray}\label{d1int}
\mathcal{D}^{(\textrm{int})}_{1,\sigma}(t)&\!\!=\!\!&\frac{t}{48\Omega^{(\textrm{int})}_{0,\sigma}(t)}
\Big\{3 \left[\left(\gm^2-12\right) \sigma _-+\left(4-3 \gm^2\right) \sigma _+-16 \gm mR\right]\nonumber\\
&&-3\sigma\left(\gm^2-4\right)t
+\left[\left(28-5 \gm^2\right) \sigma _-+\left(7 \gm^2-20\right) \sigma _+\right]t^{2}+5\sigma \left(\gm^2-4\right)t^{3}\Big\}\;,
\nonumber\\[1.5ex] \label{d2int}
\mathcal{D}^{(\textrm{int})}_{2,\sigma}(t)&\!\!=\!\!&\frac{t^{2}}{4608 (\Omega^{(\textrm{int})}_{0,\sigma}(t))^{2}} 
\Big\{36 \Big[-64 \gm^2 (mR)^2+\left(\gm^2-4\right) \sigma _- \left(\gm^2+16 \gm mR+12\right)\nonumber\\
&&-\left(\gm^2-4\right) \sigma _+ \left(3 \gm^2+16 \gm mR+4\right)\Big]+72 \left(\gm^2+16 \gm mR+4\right) \left(\gm^2 \sigma _-+4 \sigma _+\right)t\nonumber\\
&&-36 \Big[\sigma _- \left(5 \gm^4-16 \left(\gm^2+4\right) \gm mR+8 \gm^2-176\right)\nonumber\\
&&+\sigma _+ \left(-11 \gm^4-16 \left(\gm^2+4\right) \gm mR+8 \gm^2+80\right)\Big]t^{2}\nonumber\\
&&-144 \left[\gm^2 \left(3 \gm^2-8\right) \sigma _--8 \left(\gm^2-6\right) \sigma _+\right]t^{3}\nonumber\\
&&-36 \left[\left(\gm^4-40 \gm^2+208\right) \sigma _-+\left(13 \gm^4-40 \gm^2+16\right) \sigma _+\right]t^{4}\nonumber\\
&&+360 \left(\gm^2-4\right) \left(\gm^2 \sigma _--4 \sigma _+\right)t^{5}+180 \left(\gm^2-4\right)^{2}t^{6}\Big\}\;,
\\[1.5ex] \label{d3int}
\mathcal{D}^{(\textrm{int})}_{3,\sigma}(t)&\!\!=\!\!&\frac{t^{3}}{1658880(\Omega^{(\textrm{int})}_{0,\sigma}(t))^{3}}\sum_{i=0}^{9}p^{(\textrm{int})}_{i,\sigma}(\gm,mR)t^{i}\;,
\end{eqnarray}}
\end{small}
where
\begin{small}
\begin{eqnarray} 
p^{(\textrm{int})}_{0,\sigma}(\gm,mR)&\!\!\!=\!\!\!&135 \Big[-4096 \gm^3 (mR)^3+\nonumber\\
&&\sigma _- \Big(25\gm^6+204 \gm^4+1536 \left(\gm^2-4\right) \gm^2 (mR)^2+432 \gm^2\nonumber\\
&&+96 \left(\gm^4+24 \gm^2-48\right) \gm mR-4032\Big)+\sigma _+ \Big(-63 \gm^6+108 \gm^4\\
&&-1536 \left(\gm^2-4\right) \gm^2 (mR)^2+816 \gm^2-96 \left(3 \gm^4-24 \gm^2-16\right) \gm mR+1600\Big)\Big]\;,\nonumber
\\[1.5ex]
p^{(\textrm{int})}_{1,\sigma}(\gm,mR)&\!\!=\!\!&405\Big[\sigma _- \Big(25 \gm^6+1024 \gm^4 (mR)^2+68 \gm^4+112 \gm^2\nonumber\\
&&+32 \left(3 \gm^4+48 \gm^2+16\right) \gm mR+960\Big)
+\sigma _+ \Big(15 \gm^6+28 \gm^4+4096 \gm^2 (mR)^2\nonumber\\
&&+272 \gm^2+32 \left(\gm^4+48 \gm^2+48\right) \gm mR+1600\Big)\Big]\;,
\end{eqnarray}
\begin{eqnarray}
	p^{(\textrm{int})}_{2,\sigma}(\gm,mR)&\!\!=\!\!&-162 \Big[\sigma _- \Big(203 \gm^6+160 \left(\gm^4-80\right) \gm mR\nonumber\\
	&&+1756 \gm^4-1280 \left(\gm^2+4\right) \gm^2 (mR)^2-3536 \gm^2-29248\Big)
	-\sigma _+ \Big(457 \gm^6\\
	&&+160 \left(5 \gm^4-16\right) \gm mR+884 \gm^4+1280 \left(\gm^2+4\right) \gm^2 (mR)^2-7024 \gm^2-12992\Big)\Big]\;,\nonumber
\\[1.5ex]
p^{(\textrm{int})}_{3,\sigma}(\gm,mR)&\!\!=\!\!&-54 \Big[\sigma _- \Big(2327 \gm^6+1908 \gm^4-33552 \gm^2\nonumber\\
&&+480 \left(7 \gm^4+24 \gm^2+48\right) \gm mR+77632\Big)
+\sigma _+ \Big(1213 \gm^6-8388 \gm^4+7632 \gm^2\nonumber\\
&&+480 \left(3 \gm^4+24 \gm^2+112\right) \gm mR+148928\Big)\Big]\;,
\\[1.5ex]
p^{(\textrm{int})}_{4,\sigma}(\gm,mR)&\!\!=\!\!&-648 \Big[\sigma _- \left(61 \gm^6-1634 \gm^4+3144 \gm^2+20 \left(15 \gm^4+24 \gm^2+112\right) \gm mR+17024\right)\nonumber\\
&&+2 \sigma _+ \left(133 \gm^6+393 \gm^4-3268 \gm^2+10 \left(7 \gm^4+24 \gm^2+240\right) \gm mR+1952\right)\Big]\;,
\\[1.5ex]
p^{(\textrm{int})}_{5,\sigma}(\gm,mR)&\!\!=\!\!&648 \Big[2 \sigma _- \left(174 \gm^6-50 \left(\gm^4-16\right) \gm mR+237 \gm^4-4748 \gm^2+7904\right)\nonumber\\
&&+\sigma _+ \left(247 \gm^6+100 \left(\gm^4-16\right) \gm mR-2374 \gm^4+1896 \gm^2+22272\right)\Big]\;,
\\[1.5ex]
p^{(\textrm{int})}_{6,\sigma}(\gm,mR)&\!\!=\!\!&90 \Big[\left(2431 \gm^6-19860 \gm^4+23664 \gm^2+113344\right) \sigma _-\nonumber\\
&&+\left(1771 \gm^6+5916 \gm^4-79440 \gm^2+155584\right) \sigma _+\Big]\;,
\\[1.5ex]
p^{(\textrm{int})}_{7,\sigma}(\gm,mR)&\!\!=\!\!&-3510 \left(\gm^2-4\right) \Big[\left(17 \gm^4+304 \gm^2-688\right) \sigma _-+\left(43 \gm^4-304 \gm^2-272\right) \sigma _+\Big]\;,
\\[1.5ex]
p^{(\textrm{int})}_{8,\sigma}(\gm,mR)&\!\!=\!\!&-675 \left[\gm^2-4\right)^2 \left(\left(221 \gm^2+308\right) \sigma _-+\left(77 \gm^2+884\right) \sigma _+\right]\;,
\\[1.5ex]
p^{(\textrm{int})}_{9,\sigma}(\gm,mR)&\!\!=\!\!&49725\sigma \left(\gm^2-4\right)^3\;.
\end{eqnarray}
\end{small}
\\
For the exterior region we find
\begin{small}
\begin{eqnarray}\label{d1ext}
\mathcal{D}^{(\textrm{ext})}_{1,\sigma}(t)&=&\frac{t}{48\Omega^{(\textrm{ext})}_{0,\sigma}(t)}
\Big\{-3 \left[\left(\gm^2-12\right) \sigma _-+\left(4-3 \gm^2\right) \sigma _++16 \gm mR\right]\nonumber\\
&-&3\sigma \left(\gm^2-4\right)t+\left[\left(5 \gm^2-28\right) \sigma _-+\left(20-7 \gm^2\right) \sigma _+\right]t^{2}+5\sigma \left(\gm^2-4\right)t^{3}
\Big\}\;,
\end{eqnarray}
\begin{eqnarray}\label{d2ext}
\mathcal{D}^{(\textrm{ext})}_{2,\sigma}(t)&\!\!=\!\!&\frac{t^{2}}{4608 (\Omega^{(\textrm{ext})}_{0,\sigma}(t))^{2}} 
\Bigg\{36 \Big[-64 \gm^2 (mR)^2+\left(\gm^2-4\right) \sigma _- \left(\gm^2-16 \gm mR+12\right)\nonumber\\
&&-\left(\gm^2-4\right) \sigma _+ \left(3 \gm^2-16 \gm mR+4\right)\Big]
-72 \left(\gm^2-16 \gm mR+4\right) \left(\gm^2 \sigma _-+4 \sigma _+\right)t\nonumber\\
&&-36 \Big[\sigma _- \left(5 \gm^4+16 \left(\gm^2+4\right) \gm mR+8 \gm^2-176\right)\nonumber\\
&&+\sigma _+ \left(-11 \gm^4+16 \left(\gm^2+4\right) \gm mR+8 \gm^2+80\right)\Big]t^{2}\nonumber\\
&&+144 \left[\gm^2 \left(3 \gm^2-8\right) \sigma _--8 \left(\gm^2-6\right) \sigma _+\right]t^{3}\nonumber\\
&&-36 \left[\left(\gm^4-40 \gm^2+208\right) \sigma _-+\left(13 \gm^4-40 \gm^2+16\right) \sigma _+\right]t^{4}\nonumber\\
&&-360 \left(\gm^2-4\right) \left(\gm^2 \sigma _--4 \sigma _+\right)t^{5}+
180 \left(\gm^2-4\right)^2t^{6}\Bigg\}\;,
\\[1.5ex]
\label{d3ext}
\mathcal{D}^{(\textrm{ext})}_{3,\sigma}(t)&\!\!=\!\!&\frac{t^{3}}{1658880(\Omega^{(\textrm{ext})}_{0,\sigma}(t))^{3}}\sum_{i=0}^{9}p^{(\textrm{ext})}_{i,\sigma}(\gm,mR)t^{i}\;,
\end{eqnarray}
\end{small}
where 
\begin{small}
\begin{eqnarray} 
p^{(\textrm{ext})}_{0,\sigma}(\gm,mR)&\!\!=\!\!&-135 \Big[4096 \gm^3 (mR)^3\nonumber\\
&&+\sigma _- \Big(25 \gm^6+204 \gm^4+1536 (\gm-2) (\gm+2) \gm^2 (mR)^2+432 \gm^2\nonumber\\
&&-96 \left(\gm^4+24 \gm^2-48\right) \gm mR-4032\Big)+\sigma _+ \Big(-63 \gm^6+108 \gm^4\nonumber\\
&&-1536 (\gm^{2}-4)\gm^2 (mR)^2+816 \gm^2+96 \left(3 \gm^4-24 \gm^2-16\right) \gm mR+1600\Big)\Big]\;,
\\[1.5ex] 
p^{(\textrm{ext})}_{1,\sigma}(\gm,mR)&\!\!=\!\!&405 \Big[\sigma _- \Big(25 \gm^6+1024 \gm^4 (mR)^2+68 \gm^4+112 \gm^2\nonumber\\
&&-32 \left(3 \gm^4+48 \gm^2+16\right) \gm mR+960\Big)
+\sigma _+ \Big(15 \gm^6+28 \gm^4+4096 \gm^2 (mR)^2\nonumber\\
&&+272 \gm^2-32 \left(\gm^4+48 \gm^2+48\right) \gm mR+1600\Big)\Big]\;,
\\[1.5ex]
p^{(\textrm{ext})}_{2,\sigma}(\gm,mR)&\!\!=\!\!&162 \Big[\sigma _- \Big(203 \gm^6-160 \left(\gm^4-80\right) \gm mR+1756 \gm^4\nonumber\\
&&-1280 \left(\gm^2+4\right) \gm^2 (mR)^2-3536 \gm^2-29248\Big)-\sigma _+ \Big(457 \gm^6-160 \left(5 \gm^4-16\right) \gm mR\nonumber\\
&&+884 \gm^4+1280 \left(\gm^2+4\right) \gm^2 (mR)^2-7024 \gm^2-12992\Big)\Big]\;,
\\[1.5ex] 
p^{(\textrm{ext})}_{3,\sigma}(\gm,mR)&\!\!=\!\!&-54 \Big[\sigma _- \Big(2327 \gm^6+1908 \gm^4-33552 \gm^2\nonumber\\
&&-480 \left(7 \gm^4+24 \gm^2+48\right) \gm mR+77632\Big)+\sigma _+ \Big(1213 \gm^6-8388 \gm^4\nonumber\\
&&-480 \left(3 \left(\gm^2+8\right) \gm^2+112\right) \gm mR+7632 \gm^2+148928\Big)\Big]\;,
\end{eqnarray}
\begin{eqnarray}
p^{(\textrm{ext})}_{4,\sigma}(\gm,mR)&\!\!=\!\!&648 \Big[\sigma _- \Big(61 \gm^6-1634 \gm^4+3144 \gm^2-20 \left(15 \gm^4+24 \gm^2+112\right) \gm mR+17024\Big)\nonumber\\
&&+2 \sigma _+ \Big(133 \gm^6+393 \gm^4-3268 \gm^2-10 \left(7 \gm^4+24 \gm^2+240\right) \gm mR+1952\Big)\Big]\!,
\\[1.5ex]
p^{(\textrm{ext})}_{5,\sigma}(\gm,mR)&\!\!=\!\!&648 \Big[2 \sigma _- \Big(174 \gm^6+50 \left(\gm^4-16\right) \gm mR+237 \gm^4-4748 \gm^2+7904\Big)\nonumber\\
&&+\sigma _+ \Big(247 \gm^6-100 \left(\gm^4-16\right) \gm mR-2374 \gm^4+1896 \gm^2+22272\Big)\Big]\;,
\\[1.5ex]
p^{(\textrm{ext})}_{6,\sigma}(\gm,mR)&\!\!=\!\!&-90 \Big[\left(2431 \gm^6-19860 \gm^4+23664 \gm^2+113344\right) \sigma _-\nonumber\\
&&+\left(1771 \gm^6+5916 \gm^4-79440 \gm^2+155584\right) \sigma _+\Big]\;,
\\[1.5ex] 
p^{(\textrm{ext})}_{7,\sigma}(\gm,mR)&\!\!=\!\!&-3510 \left(\gm^2-4\right) \Big[\left(17 \gm^4+304 \gm^2-688\right) \sigma _-+\left(43 \gm^4-304 \gm^2-272\right) \sigma _+\Big]\;,
\\[1.5ex]
p^{(\textrm{ext})}_{8,\sigma}(\gm,mR)&\!\!=\!\!&675 \left(\gm^2-4\right)^2 \left[\left(221 \gm^2+308\right) \sigma _-+\left(77 \gm^2+884\right) \sigma _+\right]\;,
\\[1.5ex]
p^{(\textrm{ext})}_{9,\sigma}(\gm,mR)&\!\!=\!\!&49725\sigma \left(\gm^2-4\right)^3\;.
\end{eqnarray}
\end{small}

\section{Small-$\boldsymbol{m}$ expansion of the integral appearing in $\boldsymbol{b^{(\textrm{int}/\textrm{ext})}_{0}}$ and $A^{(\textrm{int}/\textrm{ext})}_{\textbf{\textit{i}}}$}\label{appb}

In this appendix we provide the small-$m$ asymptotic expansion of the integrals that 
appear in the analysis of the terms $b^{(\textrm{int}/\textrm{ext})}_{0}$ and $A^{(\textrm{int}/\textrm{ext})}_{i}$. The integrals have the general form
\begin{eqnarray}\label{integral}
I_{n}(s,\alpha)=t_{j}^{2s+\alpha+1}\int^{1}_{0}u^{2s+\alpha}(1-u^{2})^{-s}(1+\beta t_{j}u)^{-n}\du\;,
\end{eqnarray} 
with $\alpha,n\in\mathbb{N}^{+}$, and $-(\alpha+1)/2<\Re(s)<1$. The mass is contained in the 
term $t_{j}$ as follows
\begin{equation}\label{massint}
t_{j}=\left(1+\frac{m^2 R^2}{j^2}\right)^{-\frac{1}{2}}\;.
\end{equation}
There are only two terms in (\ref{integral}) that depend on $m$. The small-$m$ expansion
of $t_{j}^{2s+\alpha+1}$ can be obtained by using the binomial series, while the small-$m$
expansion of $(1+\beta t_{j}u)^{-n}$ can be computed by utilizing the Taylor expansion for the composition of two functions
\begin{equation}
f(g(\nu))=f(g(0))+\sum_{k=1}^{\infty}\frac{\nu^{k}}{k!}\left(\frac{\textrm{d}^{k}}{\textrm{d}\nu^{k}}f(g(\nu))\right)\Bigg|_{\nu=0}\;,
\end{equation}
and then by writing the higher derivatives of the composition in terms of
Fa\`a di Bruno's formula
\begin{equation}
\left(\frac{\textrm{d}^{k}}{\textrm{d}\nu^{k}}f(g(\nu))\right)\Bigg|_{\nu=0}=\sum_{l=0}^{k}f^{(l)}(g(0))
B_{k,l}\left(g'(0),g''(0),\ldots,g^{(k-l+1)}(0)\right)\;,
\end{equation}  
where $B_{k,l}\left(x_1,x_2,\ldots,x_{k-l+1}\right)$ are the Bell polynomials \cite{bell1934}. These last remarks allow us to write
\begin{equation}\label{exp1}
t_{j}^{2s+\alpha+1}(1+\beta t_{j}u)^{-n}=\sum_{p=0}^{\infty}C_{p}(s,\alpha,\beta u)\left(\frac{mR}{j}\right)^{2p}\;,
\end{equation}
where we have defined the functions
\begin{equation}
	\begin{aligned}
C_{p}(s,\alpha,\beta u)&=\sum_{l=0}^{p}\binom{-s-\alpha/2-1/2}{p-l}\frac{1}{l!}\sum_{r=0}^{l}
	(-1)^{r}\frac{(n+r-1)!}{(n-1)!}\\
&\times(1+\beta u)^{-n-r}(\beta u)^{r}
B_{l,r}\left[\binom{-1/2}{1}1!,\ldots,\binom{-1/2}{l-r+1}(l-r+1)!\right]\;.
	\end{aligned}
\end{equation}
By substituting the expression (\ref{exp1}) in the integral (\ref{integral}), we obtain
the following expansion
\begin{eqnarray}\label{exp2}
I_{n}(s,\alpha)&=&\sum_{p=0}^{\infty}\left(\frac{mR}{j}\right)^{2p}\sum_{l=0}^{p}\sum_{r=0}^{l}\binom{-s-\alpha/2-1/2}{p-l}\frac{(-1)^{r}\beta^{r}}{l!}\frac{(n+r-1)!}{(n-1)!}\nonumber\\
&\times&B_{l,r}\left[\binom{-1/2}{1}1!,\ldots,\binom{-1/2}{l-r+1}(l-r+1)!\right]\nonumber\\
&\times&\int_{0}^{1}u^{2s+\alpha+r}(1-u^{2})^{-s}(1+\beta u)^{-n-r}\du\;.
\end{eqnarray}
The integral in (\ref{exp2}) can be solved in closed form in terms of the first Appell hypergeometric function. In fact, by using 3.211 in \cite{gradshteyn2007} one finds
\begin{equation*}
	\begin{aligned}
\int_{0}^{1}u^{2s+\alpha+r}(1-u^{2})^{-s}(1+\beta u)^{-n-r}\du&=\frac{\Gamma(-s+1)\Gamma(2s+\alpha+r+1)}{\Gamma(s+\alpha+r+2)}\\
&\times F_{1}(2s+\alpha+r+1,s,n+r,s+\alpha+r+2;-1,-\beta)\;.
	\end{aligned}
\end{equation*}
By substituting this explicit expression for the integral in (\ref{exp2}) we can finally write
\begin{equation}\label{exp3}
I_{n}(s,\alpha)=\sum_{p=0}^{\infty}\Lambda_{p}(s,\alpha,n,\beta)\left(\frac{mR}{j}\right)^{2p}\;,
\end{equation}
where
\begin{eqnarray}
\Lambda_{p}(s,\alpha,n,\beta)&\!\!=\!\!&\sum_{l=0}^{p}\sum_{r=0}^{l}\binom{-s-\alpha/2-1/2}{p-l}\frac{(-1)^{r}\beta^{r}}{l!}\frac{(n+r-1)!}{(n-1)!}\frac{\Gamma(-s+1)\Gamma(2s+\alpha+r+1)}{\Gamma(s+\alpha+r+2)}\nonumber\\
&\times&B_{l,r}\left[\binom{-1/2}{1}1!,\ldots,\binom{-1/2}{l-r+1}(l-r+1)!\right]\nonumber\\
&\times& F_{1}(2s+\alpha+r+1,s,n+r,s+\alpha+r+2;-1,-\beta)\;.
\end{eqnarray}

\bibliography{Bibliography}{}
\bibliographystyle{unsrt}

\end{document}